\documentclass[12pt,a4paper]{article}
\pdfoutput=1
\usepackage[utf8]{inputenc}
\usepackage{epsf,amsmath,amssymb,graphicx,dcolumn}
\usepackage{caption}
\usepackage{subcaption}
\usepackage{scalefnt,ulem,pstricks}
\usepackage{booktabs,multirow,tabularx}
\usepackage[colorlinks=true,allcolors={blue!70!black}]{hyperref}
\usepackage{cleveref}
\usepackage{color}
\usepackage{rotating}
\usepackage{microtype}
\usepackage[titletoc,title]{appendix}
\usepackage[numbers,sort&compress]{natbib}
\usepackage{amsmath,amsfonts,amsthm,bm}
\providecommand{\href}[2]{#2}

\newcommand\as{\alpha_{\mathrm{S}}}

\def\to{\rightarrow}

\newcommand\Matrix{{\sc Matrix}}
\newcommand\Munich{{\sc Munich}}
\newcommand\OpenLoops{{\sc OpenLoops}}

\newcommand{\qt}{\ensuremath{q_T}}
\newcommand{\pt}{\ensuremath{p_T}}

\newcommand{\eqn}[1]{Eq.\,(\ref{#1})}

\newcommand{\fig}[1]{Figure~\ref{#1}}
\newcommand{\figs}[1]{Figures~\ref{#1}}
\newcommand{\tab}[1]{Table~\ref{#1}}

\def\refta#1{\mbox{Table~\ref{#1}}}

\def\citere#1{\mbox{Ref.~\cite{#1}}}
\def\citeres#1{\mbox{Refs.~\cite{#1}}}

\newcommand{\rcut}{\ensuremath{r_{\mathrm{cut}}}}

\newcommand{\CF}{C_{\mathrm{F}}}

\newcommand{\NC}{N_{\mathrm{c}}}
\newcommand{\NF}{N_{\mathrm{f}}}
\newcommand{\TF}{T_{\mathrm{F}}}

\newcommand{\mggg}{\ensuremath{m_{\gamma\gamma\gamma}}}
\newcommand{\ptggg}{\ensuremath{p_{T,\gamma\gamma\gamma}}}

\newcommand{\ptgi}{\ensuremath{p_{T,\gamma_i}}}
\newcommand{\ptgone}{\ensuremath{p_{T,\gamma_1}}}
\newcommand{\ptgtwo}{\ensuremath{p_{T,\gamma_2}}}
\newcommand{\ptgthree}{\ensuremath{p_{T,\gamma_3}}}
\newcommand{\ptgonetwo}{\ensuremath{p_{T,\gamma_1\gamma_2}}}
\newcommand{\ptgonethree}{\ensuremath{p_{T,\gamma_1\gamma_3}}}
\newcommand{\ptgtwothree}{\ensuremath{p_{T,\gamma_2\gamma_3}}}

\newcommand{\dRgg}{\ensuremath{\Delta R_{\gamma\gamma}}}
\newcommand{\etag}{\ensuremath{\eta_{\gamma}}}

\usepackage{etoolbox}
\makeatletter
\patchcmd{\@sect}{#8}{\boldmath #8}{}{}
\let\ori@chapter\@chapter
\def\@chapter[#1]#2{\ori@chapter[\boldmath#1]{\boldmath#2}}
\makeatother

\begin{document} 
\begin{flushright}
\vspace*{-1.5cm}
MPP-2020-173\
\end{flushright}
\vspace{0.cm}

\begin{center}
{\Large \bf Triphoton production at hadron colliders in NNLO QCD}
\end{center}

\begin{center}
{\bf Stefan Kallweit$^{(a)}$}, {\bf Vasily Sotnikov$^{(b)}$}, and {\bf Marius Wiesemann$^{(b)}$}

$^{(a)}$ Dipartimento di Fisica G. Occhialini, Universit\`a degli Studi di Milano-Bicocca and INFN, Piazza della Scienza 3, 20126 Milano, Italy\\
$^{(b)}$ Max-Planck-Institut f\"ur Physik, F\"ohringer Ring 6, 80805 M\"unchen, Germany

\href{mailto:stefan.kallweit@cern.ch}{\tt stefan.kallweit@cern.ch}\\
\href{mailto:sotnikov@mpp.mpg.de}{\tt sotnikov@mpp.mpg.de}\\
\href{mailto:marius.wiesemann@cern.ch}{\tt marius.wiesemann@cern.ch}

\end{center}

\begin{center} {\bf Abstract} \end{center}\vspace{-1cm}
\begin{quote}
\pretolerance 10000

We present next-to-next-to-leading-order~(NNLO) QCD corrections to the 
production of three isolated photons in hadronic collisions at the fully differential level.
We employ \qt{} subtraction within \Matrix{} and an efficient implementation of analytic two-loop amplitudes 
in the leading-colour approximation
to achieve the first on-the-fly calculation for this process at NNLO accuracy.
Numerical results are presented for proton--proton collisions at energies ranging from 7\,TeV to 100\,TeV.
We find full agreement with the 8\,TeV results of \citere{Chawdhry:2019bji} and confirm that NNLO 
corrections are indispensable to describe ATLAS 8\,TeV data. 
In addition, we demonstrate the significance of NNLO corrections for future precision studies of triphoton 
production at higher collision energies.
\end{quote}

\parskip = 1.2ex

Precision studies have become of major importance in the rich physics programme at the Large Hadron Collider (LHC).
Many LHC reactions, in particular $2\to 1$ and $2\to 2$ processes, are not only measured, but
also predicted at a remarkable accuracy by now. Prime examples are colour singlet processes,
such as vector-boson pair production, cf. for instance the recent $Z\gamma$~\cite{Aad:2019gpq} and $ZZ$~\cite{Sirunyan:2020pub} measurements 
that use the full Run-2 data. On the theoretical side next-to-next-to-leading order~(NNLO) corrections in
QCD perturbation theory are the standard now for colour singlet production involving up to two
bosons~\cite{Ferrera:2011bk,Ferrera:2014lca,Ferrera:2017zex,Campbell:2016jau,Harlander:2003ai,Harlander:2010cz,Harlander:2011fx,Buehler:2012cu,Marzani:2008az,Harlander:2009mq,Harlander:2009my,Pak:2009dg,Neumann:2014nha,Catani:2011qz,Campbell:2016yrh,Grazzini:2013bna,Grazzini:2015nwa,Campbell:2017aul,Gehrmann:2020oec,Cascioli:2014yka,Grazzini:2015hta,Heinrich:2017bvg,Kallweit:2018nyv,Gehrmann:2014fva,Grazzini:2016ctr,Grazzini:2016swo,Grazzini:2017ckn,deFlorian:2013jea,deFlorian:2016uhr,Grazzini:2018bsd,Baglio:2012np,Li:2016nrr,deFlorian:2019app}. 
In contrast, the production of three vector bosons is much more involved, both 
in terms of their measurement and in terms of their theoretical description.

Triphoton production has the largest cross section among the triboson processes. Although its measured fiducial rate is comparable to 
(even slightly higher than) that of $ZZ$ production, which has been extensively 
measured at all LHC energies,
only a single fiducial measurement of three isolated photons exists so far~\cite{Aaboud:2017lxm}, which was done 
by ATLAS at 8\,TeV (see also \citere{Aad:2015bua} for an earlier new-physics search). 
One main complication
is to reject photons produced in pion decays, which are not part of the signature. Nevertheless, the 
triphoton process (and triboson production in general) offers an important physics case for precision 
phenomenology, especially in the discovery of new-physics phenomena 
through small deviations from Standard Model (SM) predictions. 
Apart from triple gauge couplings, which can be constrained already through diboson 
production, the production of three vector bosons gives direct access to anomalous quartic gauge couplings, 
e.g.\ the $Z\to \gamma\gamma\gamma$ decay has been constrained in \citere{Aad:2015bua}.
Furthermore, the triphoton final state is important to constrain anomalous Higgs couplings in 
rare Higgs boson decays~\cite{Achard:2004kn,Denizli:2019ijf,Denizli:2020wvn} or in the rare Higgs 
boson production process in association with a photon~\cite{Aguilar-Saavedra:2020rgo} with the 
Higgs boson decaying into a pair of photons. Moreover, triphoton production is relevant as a  background to 
the associated production of a photon with a beyond-the-SM (BSM) particle that decays into a photon pair, 
see \citeres{Zerwekh:2002ex,Toro:2012sv,Das:2015bda} for instance.

On the theoretical side $2\to 3$ reactions are the current edge for
NNLO QCD calculations, limited mostly by the complicated computation of two-loop 
corrections to five-point functions. However, for the progression of precision phenomenology at the LHC
it is indispensable to go beyond the current state-of-the-art for $2\to 3$ processes,
which is next-to-leading order~(NLO) QCD accuracy. Triphoton production is the only 
$2\to 3$ process for which NNLO corrections have been calculated~\cite{Chawdhry:2019bji}.\footnote{More precisely, this statement refers to NNLO calculations that require two-loop corrections to five-point functions.}
The NLO cross section for the production of three isolated photons had been calculated
already some time ago~\cite{Bozzi:2011en} using smooth-cone isolation, and 
also considering the fragmentation contribution at leading order~(LO) in \citere{Campbell:2014yka}.

In principle, there are two mechanisms that are relevant for the production of isolated photons: the {\it direct} production
in the hard process, which can be described perturbatively, and the production through {\it fragmentation} of a 
quark or a gluon, which is non-perturbative. Since the
latter production mechanism relies on the experimental determination of fragmentation 
functions with relatively large uncertainties, we will exploit smooth-cone isolation as suggested 
by Frixione in \citere{Frixione:1998jh} to completely remove the fragmentation component in an IR-safe manner.
This substantially simplifies theoretical calculations of processes with isolated photons beyond the LO.
The finite granularity of the calorimeter prevents the usage of
smooth-cone isolation at event reconstruction level, but requires an isolation 
with a fixed cone. 
However, smooth-cone parameters are tuned to mimic the applied fixed-cone isolation criteria. 
Those parameters are provided in the fiducial phase space definition of the experimental analyses,
and the respective uncertainties are estimated, see e.g.\ Table 2 of \citere{Aaboud:2017lxm}. A 
theoretical study on the isolation parameter dependence was performed in \citere{Catani:2018krb}. 

In this letter, we present a new calculation of the fully differential NNLO cross section
for the production of three isolated photons. We exploit the \Matrix{}\footnote{\textsc{Matrix} is the abbreviation of \textsc{Munich} Automates qT subtraction and Resummation to Integrate X-sections by M. Grazzini, S. Kallweit, M. Wiesemann. The program is available under \url{http://matrix.hepforge.org}.} framework~\cite{Grazzini:2017mhc},
using its fully general implementation of the \qt{}-subtraction formalism~\cite{Catani:2007vq}, and 
demonstrate that this NNLO approach is suitable to cope with $2\to 3$ colour singlet processes.
We achieve an efficient implementation of the two-loop helicity amplitudes
for $q\bar{q}\to\gamma\gamma\gamma$ in the leading-colour approximation, based on the analytic calculation of \citere{Abreu:2020cwb}.
This is the first time that a five-point two-loop amplitude is presented which is sufficiently fast to be calculated directly in the 
physical region during phase space integration.
Therefore, at variance with \citere{Chawdhry:2019bji} our calculation allows us to obtain NNLO corrections on the fly. 
Its implementation in the \Matrix{} framework for the first time enables a fully flexible calculation of a $2\to 3$ process at NNLO 
accuracy.\footnote{The 
implementation of NNLO corrections to triphoton production will be made publicly available with the next release of \Matrix{}. A 
preliminary version of the code is available from the authors upon request.}
We study numerical results for fiducial cross sections and distributions 
in proton--proton collisions for centre-of-mass energies ranging from 7\,TeV to 100\,TeV.

\begin{figure}[t]
\begin{center}
\begin{tabular}{ccccccc}
\includegraphics[width=.21\textwidth]{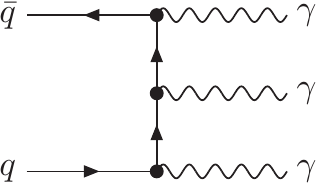} & &
\includegraphics[width=.21\textwidth]{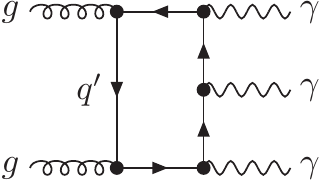} & &
\includegraphics[width=.21\textwidth]{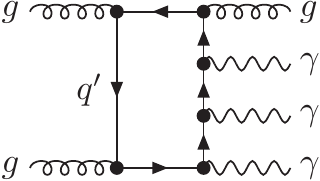} & &
\includegraphics[width=.21\textwidth]{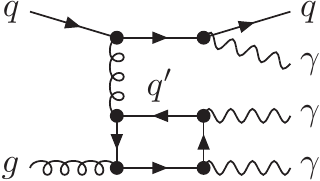} \\[1ex]
(a) & & (b) & & (c) & & (d) \\[2ex]
\end{tabular}
\end{center}\vspace{-0.4cm}
\caption{\label{fig:diag} Feynman diagrams for the production of 
three isolated photons: (a) LO diagram in the quark-annihilation channel; (b)
vanishing loop-induced diagram in the gluon-fusion channel; (c,d) first non-vanishing loop-induced contributions. Closed quark loops are included for all massive and massless flavours here, i.e. $q'=d,u,s,c,b,t$.}
\end{figure}

We consider the production of three isolated photons, i.e.\ the process 
\begin{align}
pp \rightarrow \gamma\,\gamma\,\gamma+X\nonumber\,,
\end{align}
where $X$ indicates inclusiveness over additional radiation.
At LO, the triphoton production cross section is of $\mathcal{O}(\alpha^3)$, where $\alpha$ is the electroweak (EW) coupling,
and it is $q\bar{q}$ initiated as shown in \fig{fig:diag}\,(a).
At variance with diphoton production, the loop-induced $gg$-initiated contribution shown in \fig{fig:diag}\,(b), 
which would enter at $\mathcal{O}(\alpha^3\,\as^2)$, with $\as$ being the strong coupling, vanishes 
in the case of triphoton production due to the charge-conjugation symmetry of QED$\otimes$QCD.
The first non-vanishing contributions of loop-induced type, see \fig{fig:diag}\,(c) and (d), are $gg$- or $qg$-initiated, and they
enter at $\mathcal{O}(\alpha^3\,\as^3)$, i.e.\ beyond the nominal accuracy of our calculation. 
Nevertheless, since those contributions are separately finite and represent the first non-vanishing order of the loop-induced production mode, we have calculated them
and found their effect to be below $1\%$ of the integrated NNLO cross section, and only slightly larger when considered differentially in phase space. Thus, we can safely 
neglect the loop-induced contributions in what follows.

For our calculation we employ the \Matrix{} framework~\cite{Grazzini:2017mhc}.
All tree-level and one-loop amplitudes are evaluated with \OpenLoops{}~\cite{Cascioli:2011va,Buccioni:2017yxi,Buccioni:2019sur}.
At the two-loop level we have performed a novel implementation based on the
analytic expressions in \citere{Abreu:2020cwb}, which is highly efficient and evaluates
the 2-loop hard function within few seconds for each phase space point, as discussed in more 
detail below.
NNLO accuracy is achieved by a fully general implementation of the \qt{}-subtraction formalism~\cite{Catani:2007vq}
within \Matrix{}. The NLO parts therein (for $\gamma\gamma\gamma$ and $\gamma\gamma\gamma$+$1$-jet)
are calculated by \Munich{}\footnote{The Monte
Carlo program \Munich{} --- the abbreviation stands for
``MUlti-chaNnel Integrator at Swiss (CH) precision'',  by S.~Kallweit --- features a general implementation of an
efficient, multi-channel based phase space integration and computes
both NLO QCD and NLO EW~\cite{Kallweit:2014xda,Kallweit:2015dum} corrections
to arbitrary SM processes.},
which uses the Catani--Seymour dipole subtraction method~\cite{Catani:1996jh,Catani:1996vz}.
The \Matrix{} framework features NNLO QCD corrections to a large number of colour singlet processes
at hadron colliders.\footnote{The \Matrix{} framework was recently extended and applied to
heavy-quark production processes~\cite{Catani:2019iny,Catani:2019hip,Catani:2020tko}.}
It has already been used to obtain several 
state-of-the-art NNLO QCD predictions~\cite{Grazzini:2013bna,Grazzini:2015nwa,Cascioli:2014yka,Grazzini:2015hta,Gehrmann:2014fva,Grazzini:2016ctr,Grazzini:2016swo,Grazzini:2017ckn,Kallweit:2018nyv,deFlorian:2016uhr,Grazzini:2018bsd}\footnote{It was also used in the NNLO+NNLL computation of \citere{Grazzini:2015wpa}, and in the NNLOPS computations of \citeres{Re:2018vac,Monni:2019whf,Alioli:2019qzz,Monni:2020nks}.}, and for massive diboson processes
it was extended to combine NNLO QCD with NLO EW corrections~\cite{Kallweit:2019zez} and with NLO QCD corrections to the loop-induced gluon fusion process~\cite{Grazzini:2018owa,Grazzini:2020stb}. Through the recently implemented interface~\cite{Kallweit:2020gva,Wiesemann:2020gbm} to the 
code \textsc{RadISH}~\cite{Bizon:2017rah,Monni:2019yyr} this framework now also includes the 
resummation of transverse observables such as the transverse momentum 
of the produced colour singlet final state.

\begin{figure}[t]
\begin{center}
\includegraphics[width=.55\textheight]{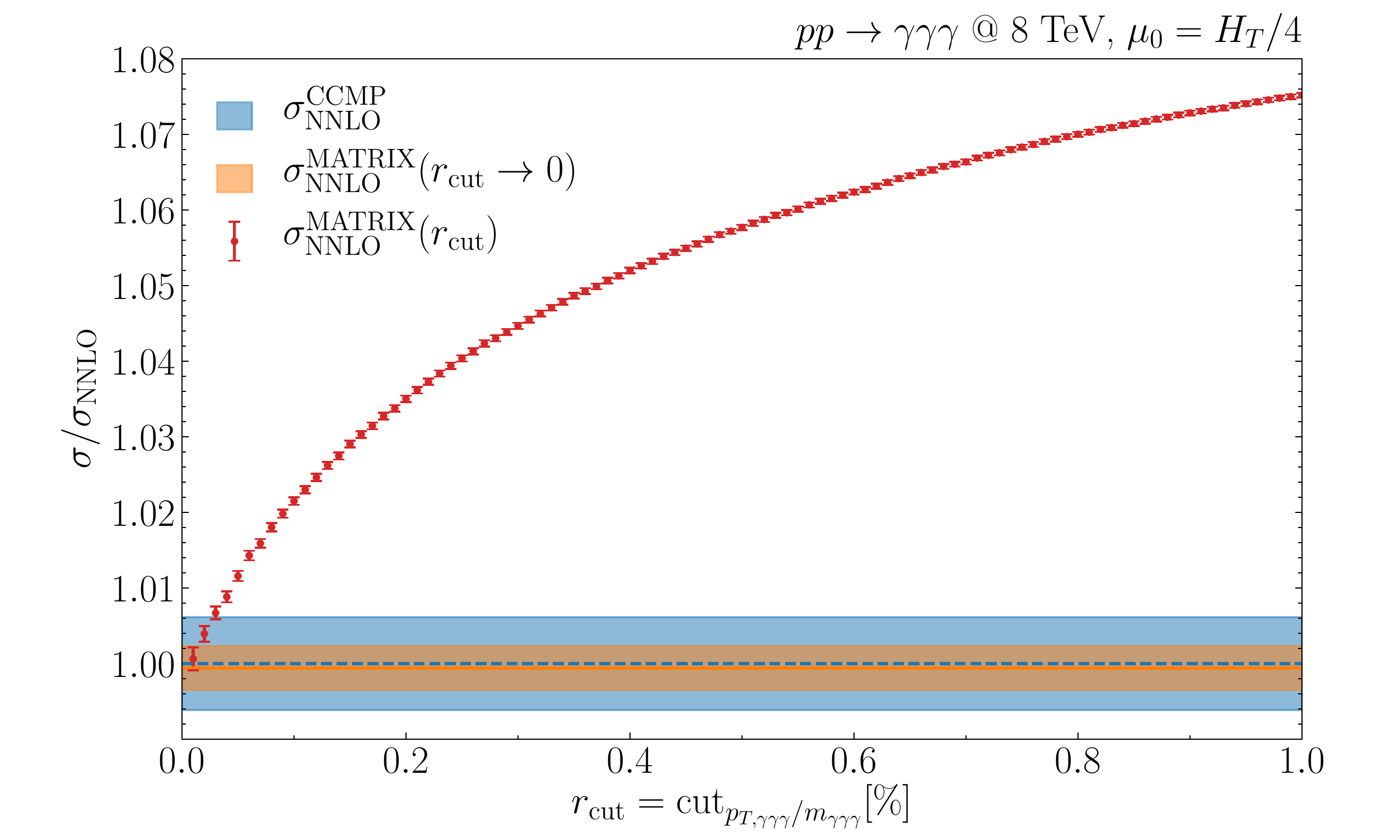}
\vspace*{1ex}
\caption{\label{fig:rcut} Dependence of the NNLO cross section for $pp\rightarrow \gamma\gamma\gamma+X$
on the slicing parameter $r_{\mathrm{cut}}$ (red points with numerical error bars),
the extrapolated cross section for $r_{\mathrm{cut}}\to0$ (orange, solid) and
comparison to the results from \citere{Chawdhry:2019bji} (blue, dashed).}
\end{center}
\end{figure}

The \qt{}-subtraction formalism is employed for the very first time for a colour singlet process
of the given complexity. Not only is triphoton production a $2\to 3$ process, it also involves 
the isolation of all of the three photons. Already processes with a single isolated photon, such as $Z\gamma$ production, 
feature rather large power corrections in the transverse momentum of the colour singlet system, 
as shown in \citere{Grazzini:2017mhc}. Due to the interplay between the smooth-cone 
isolation criteria and the slicing cutoff (introduced as $\rcut{}$ in \citere{Grazzini:2017mhc}) of the 
\qt{}-subtraction approach, 
the production of three isolated photons could be subject to relatively large systematic uncertainties at NNLO.
To deal with this issue, the slicing parameter $\rcut{}$, which is defined as a lower cut on the 
dimensionless quantity $r= \pt/m$ of the respective colour singlet, is fully monitored and controlled by \Matrix{}, 
including a completely automated cutoff extrapolation $\rcut\rightarrow 0$ performed with every run~\cite{Grazzini:2017mhc}.
 \fig{fig:rcut} shows the NNLO cross section for triphoton production, within the fiducial cuts specified in \refta{tab:cuts}, as a function of \rcut{} (which here denotes a lower cut on $r= \ptggg/\mggg$) and 
its extrapolation to $\rcut=0$ with an estimate of the respective uncertainties.
We find that \qt{} subtraction is fully capable of dealing with $2\to3$ colour singlet 
processes, even when three isolated photons are involved, and that we can control the numerical integration and the systematic uncertainties induced by the \rcut{} dependence of the cross section
at the few permille level, which fully suffices for any phenomenological application.\footnote{For the differential distributions presented in this letter 
we have performed a bin-wise extrapolation $\rcut\to 0$ as used for instance 
in \citeres{Grazzini:2017ckn,Catani:2019hip} and found extrapolation effects to depend rather mildly on the region of phase space. Therefore,
rescaling distributions computed at a fixed $\rcut$ value with the integrated cross section in the $\rcut\to 0$ limit yields a suitable approximation.}

Before presenting phenomenological results we comment in more detail on our calculation of the
double-virtual contribution.
We exploit the compact analytic expressions for the finite remainders of two-loop and one-loop helicity amplitudes reported in \citere{Abreu:2020cwb}.
They were obtained within the \texttt{C++} framework \texttt{Caravel}~\cite{Abreu:2020xvt},
based on the numerical unitarity method and analytic reconstruction techniques~\cite{Ita:2015tya,Abreu:2018zmy,Abreu:2019odu,Abreu:2018jgq,Abreu:2017hqn}.
We use the analytic expressions for the finite remainders to 
implement the numerical evaluation of the two-loop hard function, defined in Eqs.\ (12) and (62) of \citere{Catani:2013tia}, 
for the $q\bar{q}\to\gamma\gamma\gamma$ process within \Matrix{}.\footnote{To account for different definitions of the infrared subtraction operators, 
the necessary finite shifts are applied.}
The finite remainders are expressed in terms of a judiciously constructed basis of multivariate transcendental functions~\cite{Chicherin:2020oor}.
Their numerical evaluation relies on the library \texttt{PentagonFunctions++} provided in \citere{Chicherin:2020oor}.
The evaluation of the two-loop hard function on average takes only a few seconds per phase space point, and the numerical error is far below the error of the phase space integration.
Thus, the two-loop amplitude can be evaluated on the fly during phase space integration, so that there is no need to construct an interpolating function from a pre-generated 
grid in phase space for the two-loop contribution, as it was done in \citere{Chawdhry:2019bji}.
This implicates not only a vast improvement in terms of flexibility and applicability of the resulting 
NNLO calculation, but it also eliminates any interpolation uncertainties from the ensuing results.

The hard function at two-loop level for the process $q\bar{q}\to\gamma\gamma\gamma$ can be written as
\begin{multline} \label{eq:h2contributions}
  H^{(2)} = \frac{\NC^2}{4} \left( H^{(2,0)} - \frac{1}{\NC^2} (H^{(2,0)} + H^{(2,1)}) + \frac{1}{\NC^4} H^{(2,1)} \right)+ \\
    \CF{}\TF{}\NF{} \, H^{(2,\NF)}+ \CF{} \TF{} \left(\sum_{f=1}^{\NF} Q_f^2\right) \, H^{(2,\tilde{\NF})}, \qquad
\end{multline}
where $\NC$ is the number of colours, $\NF$ the number of light quark flavours, $\CF = (\NC^2-1)/(2\NC)$ the quadratic Casimir operator of the fundamental representation,
and $Q_f$ the ratio of the electric charge of the quark with flavour $f$ to the electric charge of the quarks in the initial state, and $T_F = 1/2$.
The functions $H^{(2,\NF)}$ and $H^{(2,\tilde{\NF})}$ describe the contributions from two types of Feynman diagrams with closed quark loops.
The former captures the diagrams with no photons coupling to the quark loop, and the latter those with two of the photons attached to the quark loop.\footnote{
The contribution from diagrams with one or three photons attached to the quark loop vanishes for the same reason as the amplitude $gg\to\gamma\gamma\gamma$.}
The functions $H^{(2,1)}$ and $H^{(2,\tilde{\NF})}$ require the evaluation of non-planar Feynman diagrams that are beyond reach with current computational techniques.
Thus, while our predictions include the full colour dependence in all other contributions, we compute $H^{(2)}$ in the approximation
where we keep only the leading term $\NC^2 H^{(2,0)}$ in the formal $\NC\to\infty$ limit.\footnote{The same approximation was used in \citere{Chawdhry:2019bji}.}
The contribution from $H^{(2,1)}$ is suppressed by a factor of $1/\NC^2$ and can therefore be safely neglected.
In a naive counting, the terms $H^{(2,\NF)}$ and $H^{(2,\tilde{\NF})}$ are expected to be of similar size as the leading-colour term $H^{(2,0)}$,
but for other processes, such as diphoton~\cite{Anastasiou:2002zn} or dijet~\cite{Currie:2013dwa} production for instance,
those contributions are numerically subleading with respect to the leading-colour term.
In order to gauge the quality of our approximation, we consider the directly related diphoton process as a proxy.
The two-loop hard function for this process is of the same form as given in \eqn{eq:h2contributions}, and all subleading corrections are known~\cite{Anastasiou:2002zn}.
For brevity we consider only the dominant subprocess $u\bar{u}\to \gamma\gamma$ here.
\begin{figure}[t]
\centering
\includegraphics[width=0.75\textwidth]{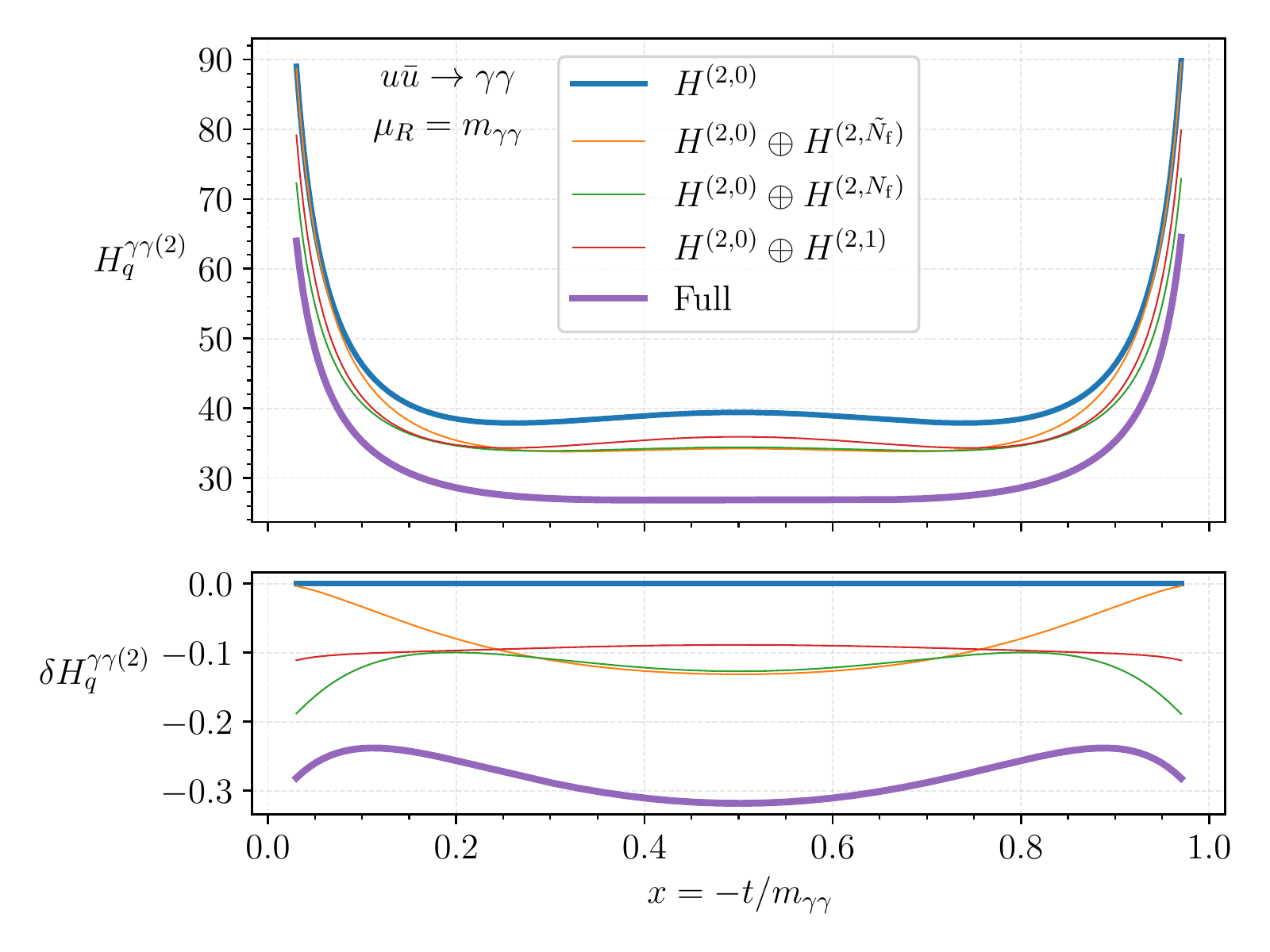}
\vspace*{1ex}
\caption{
Impact of different contributions to the two-loop hard function of the $u\bar{u}\to \gamma\gamma$ process as functions of the (dimensionless) momentum transfer.
The main frame shows the leading-colour approximation $H^{(2,0)}$, a separate curve
for the combination of $H^{(2,0)}$ with each of the three subleading corrections in \eqn{eq:h2contributions}, and the full result.
The lower frame shows the relative differences with respect to $H^{(2,0)}$.
}
\label{fig:h2contributions}
\end{figure}
\fig{fig:h2contributions} shows the impact of each individual correction beyond leading colour. 
All three corrections are of the order of $10\%$, adding up to at most $30\%$. 
In the case of triphoton production, only the functions $H^{(2,0)}$ and $H^{(2,\NF)}$ are known~\cite{Abreu:2020cwb}.
Since not all two-loop amplitudes with closed quark loops have been calculated, for consistency we do not include 
$H^{(2,\NF)}$ in our calculation.
Nevertheless, we calculated the $H^{(2,\NF)}$  contribution separately 
and found that its size is about $-18\%$ of the $H^{(2,0)}$ contribution, and largely independent 
of phase space region and centre-of-mass energy. 
This is in line with our assumption that $N_f$-related corrections are numerically subleading.
We have determined the contribution of $H^{(2)}$ in our approximation to be about $3\%$ for the integrated NNLO cross section at 8\,TeV,\footnote{
\citere{Chawdhry:2019bji} quotes a similar relative 
size of the finite part of the double-virtual corrections in their subtraction scheme for the integrated cross section. 
They further argue that even in a conservative uncertainty estimate of the approximation at hand for the two-loop amplitude, where 
one assumes it to be wrong by at most $100\%$, the 
ensuing uncertainty is still below the scale uncertainties, which should be 
sufficient for phenomenological applications. In fact, we reckon that such uncertainty 
estimate might be even too conservative as we argue in the main text.}
while it reaches up to $6\%$ in the differential distributions considered in \citere{Aaboud:2017lxm}.
We confirm this statement also for high-energy tails that are not resolved in this measurement,
and we find slightly larger effects only for the TeV ranges of some invariant-mass distributions.
The overall impact of $H^{(2)}$ on the fiducial cross section continuously decreases
with increasing collision energy, by roughly a factor of 2 when going to 100\,TeV,
and the same trend is observed for differential observables as well.
We therefore conclude that the effect of our approximation is negligible for phenomenological applications.

We present predictions for proton--proton collisions at centre-of-mass energies ($\sqrt{s}$) ranging between 7\,TeV and 100\,TeV. We employ the complex-mass scheme~\cite{Denner:2005fg} and use the $G_\mu$ scheme 
for the EW parameters, i.e.\ we evaluate the EW coupling as 
$\alpha=\sqrt{2}\,G_F \left|\mu_W^2\left(1-\mu_W^2/\mu_Z^2\right)\right|/\pi$ and the mixing angle
as $\cos\theta_W^2=\mu_W^2/\mu_Z^2$, with $\mu_V = m_V^2-i\Gamma_V\,m_V$ and $V\in\{Z,W\}$. 
The input parameters are set to the PDG~\cite{Patrignani:2016xqp} values:
$G_F = 1.16639\times 10^{-5}$\,GeV$^{-2}$, $m_W=80.385$\,GeV,
$\Gamma_W=2.0854$\,GeV, $m_Z = 91.1876$\,GeV, $\Gamma_Z=2.4952$\,GeV.
We choose the $\NF=5$ NNPDF3.1~\cite{Ball:2017nwa} sets of parton distribution functions with $\as(m_Z)=0.118$ and use the corresponding set for each perturbative order.
The central renormalization ($\mu_R$) and the factorization ($\mu_F$) scales are
set to 
\begin{align} 
\label{eq:scale}
\mu_R=\mu_F=\mu_0\equiv\,\frac14\,\left(\ptgone{}+\ptgtwo{}+\ptgthree{}\right)\,,
\end{align}
while a customary 7-point variation 
is used to estimate the uncertainties due to missing higher-order corrections.
Thus, the renormalization and factorization scales are varied
around $\mu_0$ by a factor of two with the constraint $0.5\le \mu_R/\mu_F\le 2$, and the band between
the minimum and maximum value of the cross section estimates the uncertainty. The scale setting in 
\eqn{eq:scale} was the preferred one in \citere{Chawdhry:2019bji}, which allows us to directly compare to their results.

\renewcommand{\baselinestretch}{1.5}
\begin{table}[!t]
\begin{center}
\begin{tabular}{c}
\toprule
fiducial setup for $pp\to\gamma\gamma\gamma +X$; used in the ATLAS 8\,TeV analysis of~\citere{Aaboud:2017lxm}\\
\midrule
$\ptgone\ge 27$\,GeV, \quad $\ptgtwo\ge 22$\,GeV, \quad $\ptgthree\ge 15$\,GeV, \quad$0\le | \etag | \le 1.37$ or $1.56\le |\etag|\le 2.37$\,, \\
$\dRgg\ge 0.45$, \quad $\mggg\ge 50$\,GeV, \quad Frixione isolation with $n=1$, $\delta_0=0.4$, and $E_T^{\rm ref} = 10$\,GeV\,.\\
\bottomrule
\end{tabular}
\end{center}
\renewcommand{\baselinestretch}{1.0}
\caption{\label{tab:cuts}
Definition of phase space cuts.}
\end{table}
\renewcommand{\baselinestretch}{1.0}

We study integrated cross sections and distributions in the fiducial region. 
The cuts are summarized in \tab{tab:cuts} and correspond to 
the fiducial phase space definition of the ATLAS 8\,TeV
analysis~\cite{Aaboud:2017lxm}. Those cuts involve different transverse-momentum
thresholds for the three photons as well as a requirement on their 
pseudorapidities. Furthermore, we impose a separation of each 
pair of photons in $\Delta R=\sqrt{\Delta \phi^2+\Delta\eta^2}$ and
a lower bound on the invariant mass of the three-photon system.
Finally, the photons are required to be isolated, which is achieved by 
means of Frixione's smooth-cone isolation~\cite{Frixione:1998jh} with a fixed value 
(instead of one relative to the transverse momentum of the respective photon)
of the threshold $E_T^{\rm ref}$ in the smooth-cone condition as given 
in Eq.~(3) of \citere{Grazzini:2017mhc}.

\begin{figure}[t]
\begin{center}
\begin{tabular}{cc}
\includegraphics[width=.42\textwidth]{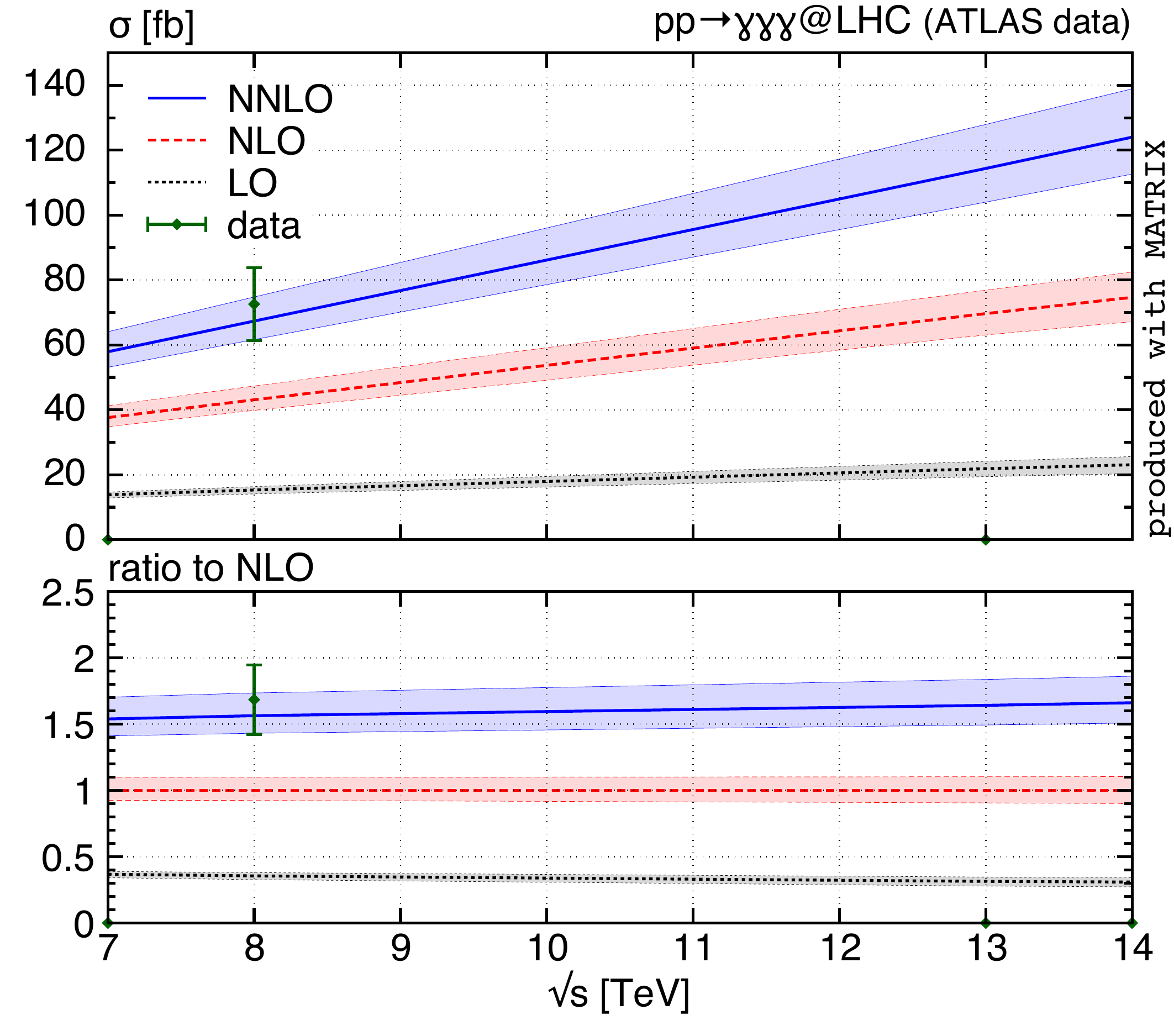} 
&
\includegraphics[width=.42\textwidth]{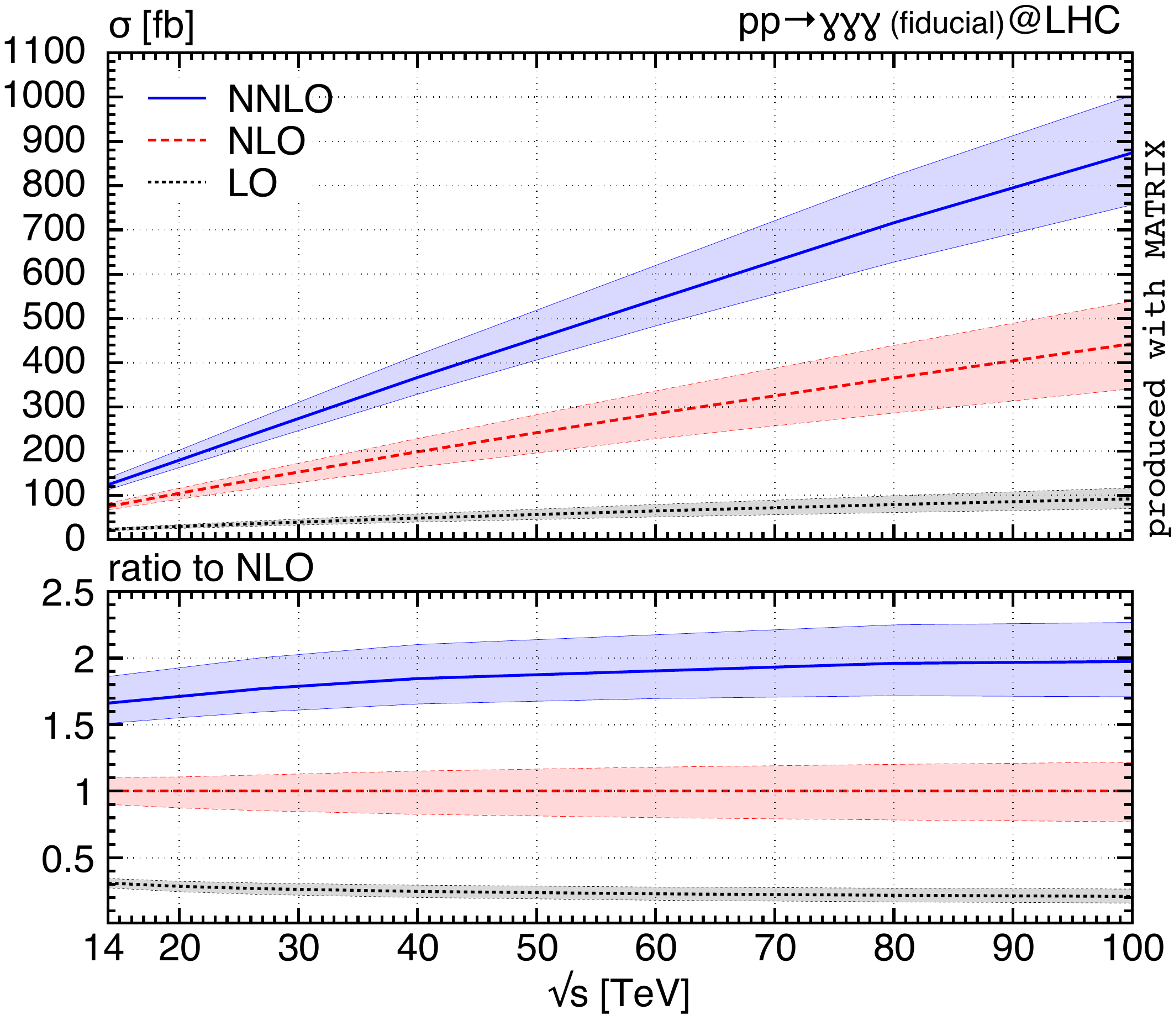}
\\
\end{tabular}
\vspace{1ex}
\caption{\label{fig:cs} Fiducial cross sections for $pp\rightarrow \gamma\gamma\gamma+X$ as a function of the centre-of-mass energy
at LO (black dotted), at NLO (red dashed), and at NNLO (blue, solid)
The green data point at 8\,TeV corresponds to the cross section 
measured by ATLAS in \citere{Aaboud:2017lxm}.}
\end{center}
\end{figure}

\begin{table}
\begin{center}
\begin{tabular}{c c c c c c c c c c c }
\toprule
$\sqrt{s}$ [TeV]
&& $\sigma_{\textrm{LO}}$ [fb]
&& $\sigma_{\textrm{NLO}}$ [fb]
&& $\sigma_{\textrm{NNLO}}$ [fb]
&& $K_{\rm NLO}$
&& $K_{\rm NNLO}$\\

\midrule

7 && $13.8237(14)_{-\phantom{0}7.0\%}^{+\phantom{0}6.0\%}$ && $\phantom{0}37.6084(35)_{-\phantom{0}7.5\%}^{+\phantom{0}9.7\%}$ && $\phantom{0}57.84(20)_{-\phantom{0}8.3\%}^{+10.7\%}$ && 2.72 && 1.54 \\[1ex] 
8 && $15.3023(15)_{-\phantom{0}8.0\%}^{+\phantom{0}6.9\%}$ && $\phantom{0}43.1076(22)_{-\phantom{0}7.6\%}^{+\phantom{0}9.9\%}$ && $\phantom{0}67.42(20)_{-\phantom{0}8.5\%}^{+11.0\%}$ && 2.82 && 1.56 \\ [1ex] 
13 && $21.8814(22)_{-11.2\%}^{+10.4\%}$ && $\phantom{0}69.6330(60)_{-\phantom{0}9.5\%}^{+10.3\%}$ && $114.60(43)_{-\phantom{0}9.1\%}^{+11.9\%}$ && 3.18 && 1.65 \\ [1ex] 
14 && $23.0839(23)_{-11.7\%}^{+10.9\%}$ && $\phantom{0}74.7875(82)_{-10.0\%}^{+10.4\%}$ && $123.83(24)_{-\phantom{0}9.2\%}^{+12.0\%}$ && 3.24 && 1.66 \\ [1ex] 
27 && $36.9540(37)_{-16.1\%}^{+16.0\%}$ && $138.797(13)\phantom{0}_{-14.8\%}^{+12.2\%}$ && $245.91(48)_{-\phantom{0}9.9\%}^{+13.2\%}$ && 3.76 && 1.77 \\ [1ex] 
100 && $92.3779(92)_{-24.0\%}^{+26.6\%}$ && $442.310(39)\phantom{0}_{-23.0\%}^{+21.7\%}$ && $878.9(24)\phantom{0}_{-13.5\%}^{+15.0\%}$ && 4.79 && 1.99 \\
 
\bottomrule

\end{tabular}

\end{center}
\renewcommand{\baselinestretch}{1.0}
\caption{\label{tab:cs} Predictions for fiducial $pp\rightarrow \gamma\gamma\gamma+X$ cross sections at different centre-of-mass energies; the numbers in brackets are integration errors, while at NNLO they also include systematic uncertainties from $\rcut{}$ dependence, see \citere{Grazzini:2017mhc}; the percentages correspond to scale uncertainties; $K_{\rm NLO}\equiv{\sigma_{\rm NLO}}/{\sigma_{\rm LO}}$\,, $K_{\rm NNLO}\equiv{\sigma_{\rm NNLO}}/{\sigma_{\rm NLO}}$\,.}
\end{table}

We start by discussing fiducial rates in proton--proton collisions as a function
of the machine energy shown in \fig{fig:cs}. The corresponding numbers 
and $K$-factors are quoted in \tab{tab:cs} at the LHC energies 7\,TeV, 8\,TeV, 13\,TeV and 14\,TeV, 
and for two potential future colliders, the HE-LHC at 27\,TeV and the FCC-hh at 100\,TeV.
We recall that producing these results is possible due to our fast and efficient calculation of triphoton 
production within \Matrix{}, which allows us to obtain predictions for any setup and any 
centre-of-mass energy within a couple of days on a medium-sized cluster. 
We find full agreement with the results in \citere{Chawdhry:2019bji}, which are restricted to 8\,TeV,
with a quoted cross section of $67.5_{-8\%}^{+11\%}$\,fb at 8\,TeV. 
We have also reproduced all differential distributions in the identical setup as 
in \citere{Chawdhry:2019bji}, and we found full agreement with the results in the ancillary files of \citere{Chawdhry:2019bji}
within their quoted statistical uncertainties of about $1\%$ ($0.1\%$) at NNLO (LO and NLO), with the following exception:
The last bins of some of their kinematical distributions were subject to a minor bug, which we could reproduce by
treating them as overflow bins, i.e.\ by including all events beyond the upper bound into the last bin,
while keeping its normalization unchanged.
\footnote{We thank Micha\l{} Czakon and Rene Poncelet for private communication and clarifications regarding that issue.}

The results in \fig{fig:cs} and \tab{tab:cs} show that higher-order corrections to triphoton production are very sizeable 
and absolutely crucial to obtain a reliable prediction. Already at LHC energies the NLO cross section is about a 
factor of three larger than the LO one, and the corrections further increase with the collision energy. Even NNLO corrections
are larger than $50$\% at the LHC and reach almost a factor of two at the FCC-hh. Thus, NNLO accuracy 
is indispensable for an accurate prediction of the triphoton cross section. Indeed, the large discrepancy of several standard deviations
observed in the ATLAS 8\,TeV measurement of \citere{Aaboud:2017lxm} when compared to NLO-accurate predictions 
 is completely eliminated by the NNLO corrections, with the NNLO cross section being fully 
consistent with the measured 8\,TeV cross section of $72.6\;^{+6.5}_{-6.5 }{\rm (stat)}\;^{+9.2}_{-9.2}{\rm (syst)}$\,fb.

Due to these sizeable corrections, one may question the validity of scale variations to estimate
uncertainties from missing higher orders. Clearly, neither the scale uncertainties 
at LO nor the ones at NLO capture the corrections of the next order.
In addition, contrary to what one usually expects, the uncertainties do not (or hardly) decrease upon inclusion of both NLO and 
NNLO corrections, especially at lower collider energies, where they even slightly increase in some cases.
The fact that scale uncertainties at LO and NLO do not cover 
higher-order corrections 
can be largely attributed to a well-known observation that opening of new partonic channels induces sizeable corrections in fixed-order predictions,
especially if the opened channels are enhanced by large initial-state flux.%
\footnote{This was discussed in detail in \citere{Chawdhry:2019bji} for triphoton production, following the corresponding considerations for diphoton production in \citere{Catani:2018krb}.}
Indeed, other processes with similarly large NLO and NNLO 
corrections, such as single Higgs production,
receive a rather small contribution at N$^3$LO, well within the uncertainties estimated from NNLO scale variations.
This argument is also supported by an observation made in \citere{Anger:2017glm} about the $pp\to Wb\bar{b}$ process which suffers from large NLO $K$-factors.
Considering the same process with a large enough number of additional jets in the final state,
such that all partonic channels are included already at LO, one observes that the $K$-factors shrink significantly, and the NLO predictions end up being within the LO scale uncertainties.
For triphoton production all partonic initial states are first included at NNLO.
Therefore, corrections beyond NNLO are expected to be smaller and presumably 
at the level of the estimated uncertainty.
Also the significant decrease from relative NLO to NNLO corrections can be seen
as a heuristic hint towards perturbative convergence.
We note that it would be interesting to investigate the impact of scale choices and
photon isolation criteria on the theoretical predictions and the respective uncertainties
in the future, as done recently in the case of diphoton production~\cite{Gehrmann:2020oec}.
A more conservative uncertainty estimate that is reliable also at LO and NLO can be obtained
by following the approach suggested in \citere{Bonvini:2020xeo}. 
Unfortunately, a systematic application of this approach to the cases where the factorization and renormalization scales are dynamically defined is not straightforward.%
\footnote{We thank Marco Bonvini for correspondence regarding this issue.}
We therefore leave studies of a probabilistic definition of theoretical uncertainties for future work.

\begin{figure}[t!]
\begin{center}
\begin{tabular}{cc}
\includegraphics[width=.42\textwidth]{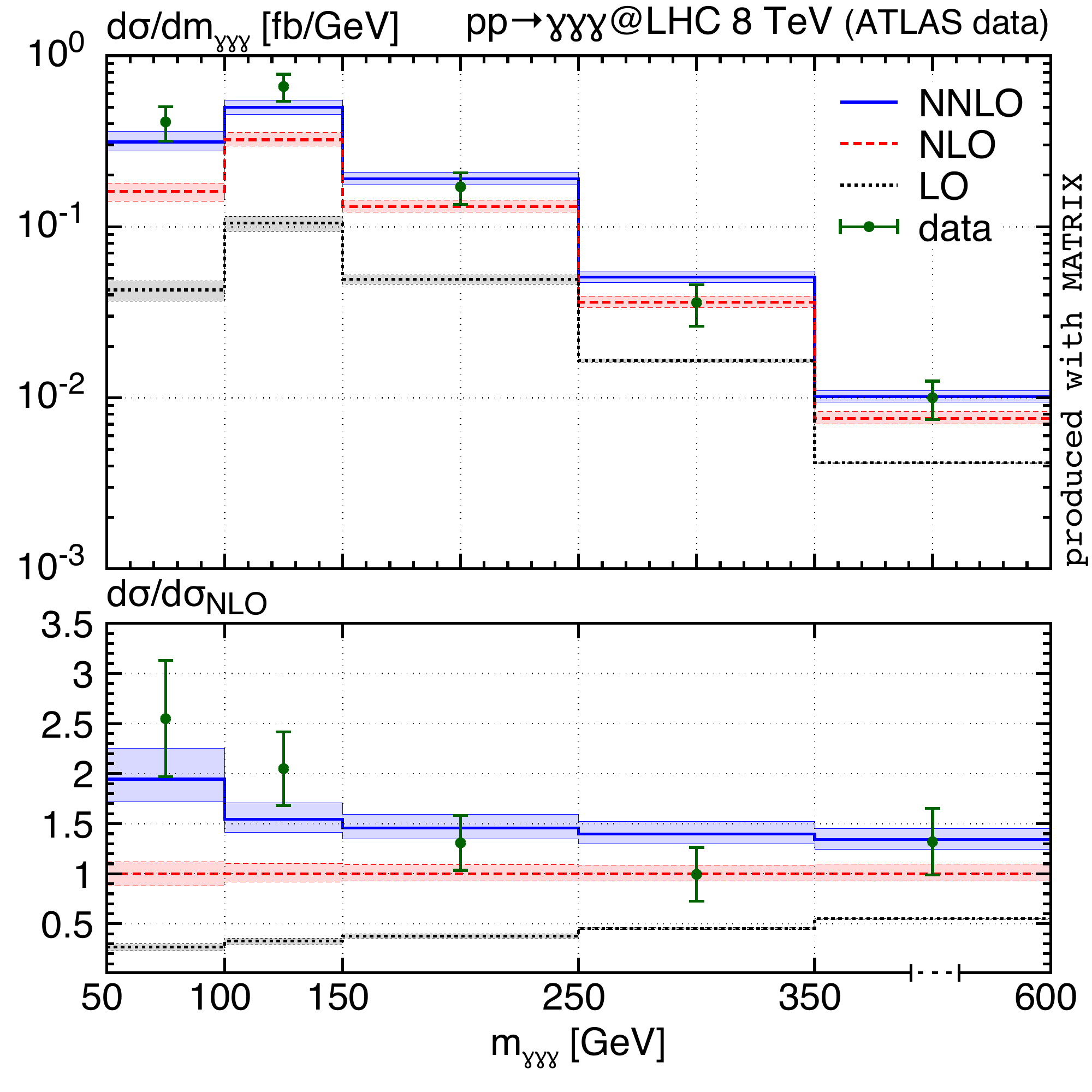} 
&
\includegraphics[width=.42\textwidth]{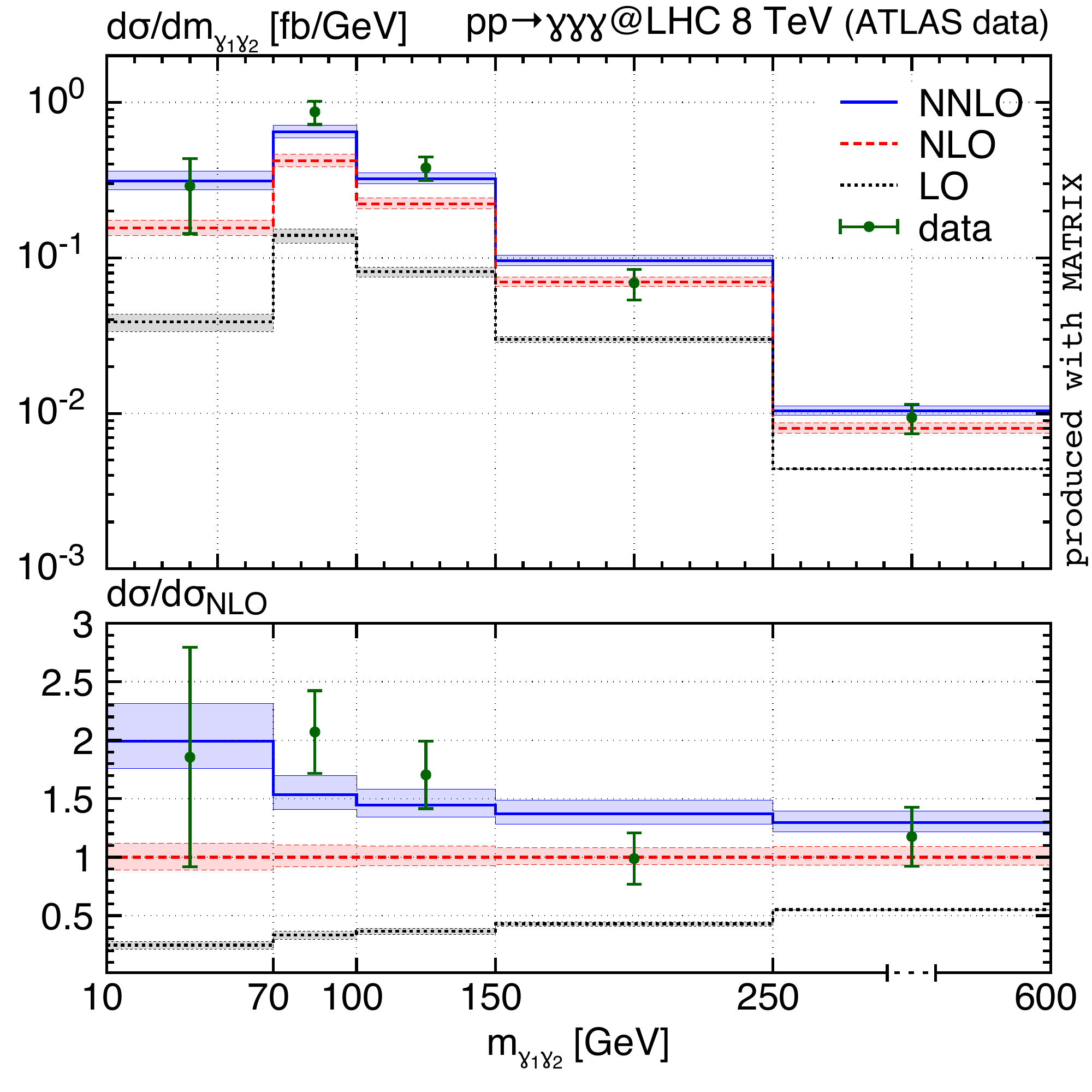}
\end{tabular}
\begin{tabular}{cc}
\includegraphics[width=.42\textwidth]{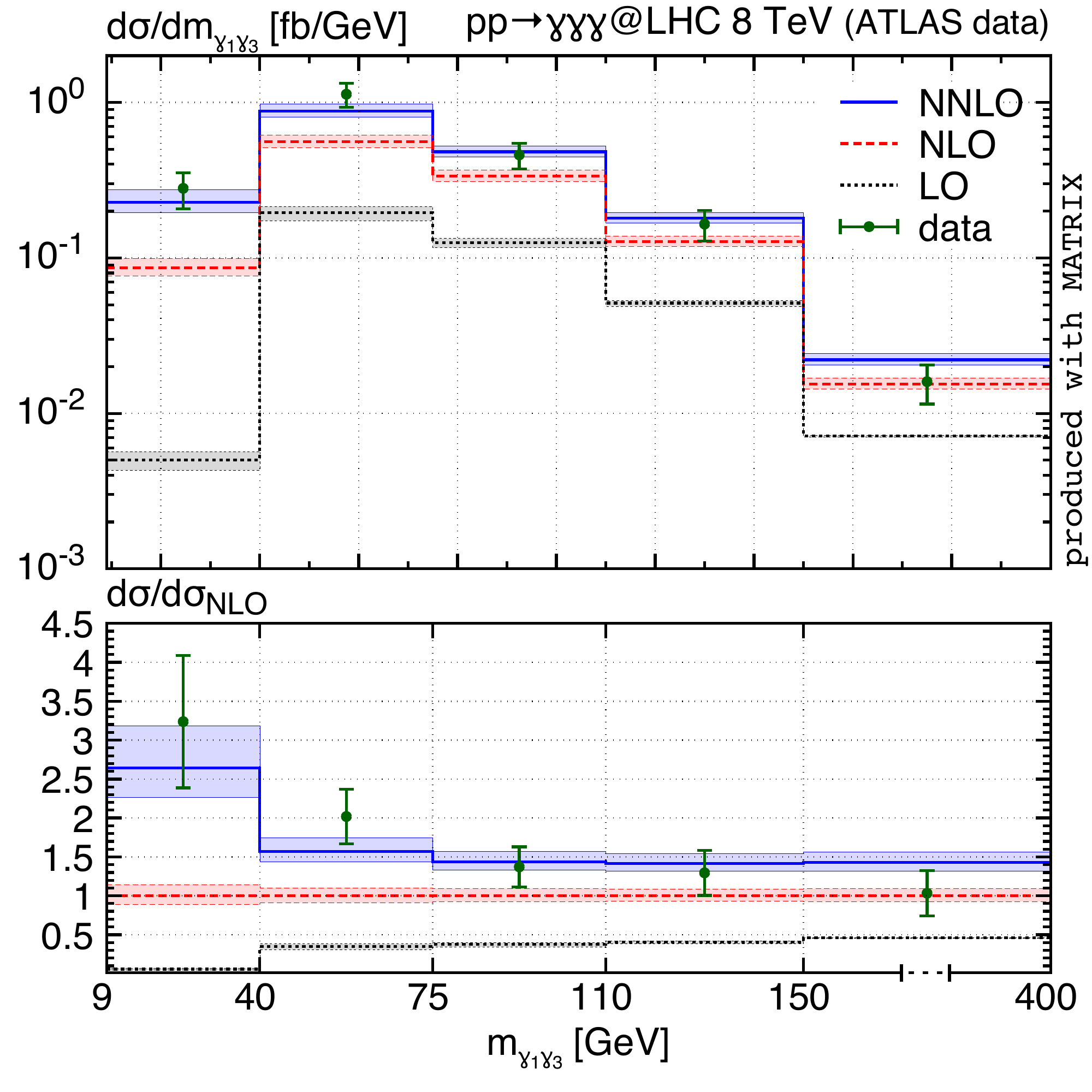}
&
\includegraphics[width=.42\textwidth]{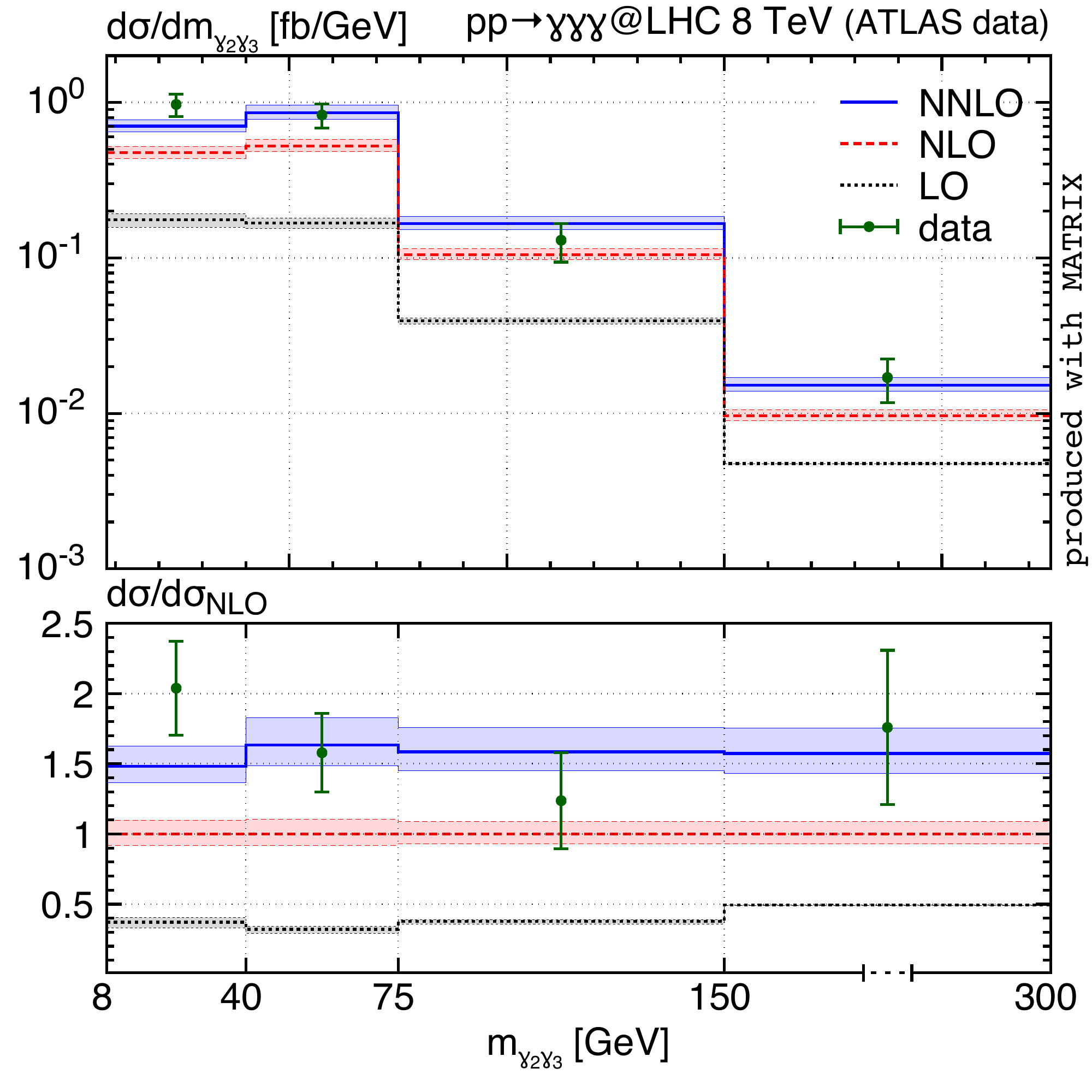}
\end{tabular}
\vspace*{1ex}
\caption{\label{fig:invmass}  Invariant-mass distribution of the 
three-photon system (top left plot) and of each photon pair compared
to 8\,TeV ATLAS data~\cite{Aaboud:2017lxm}. The colour coding corresponds to
\fig{fig:cs}.}
\end{center}
\end{figure}

\begin{figure}[t!]
\begin{center}
\begin{tabular}{cc}
\includegraphics[width=.42\textwidth]{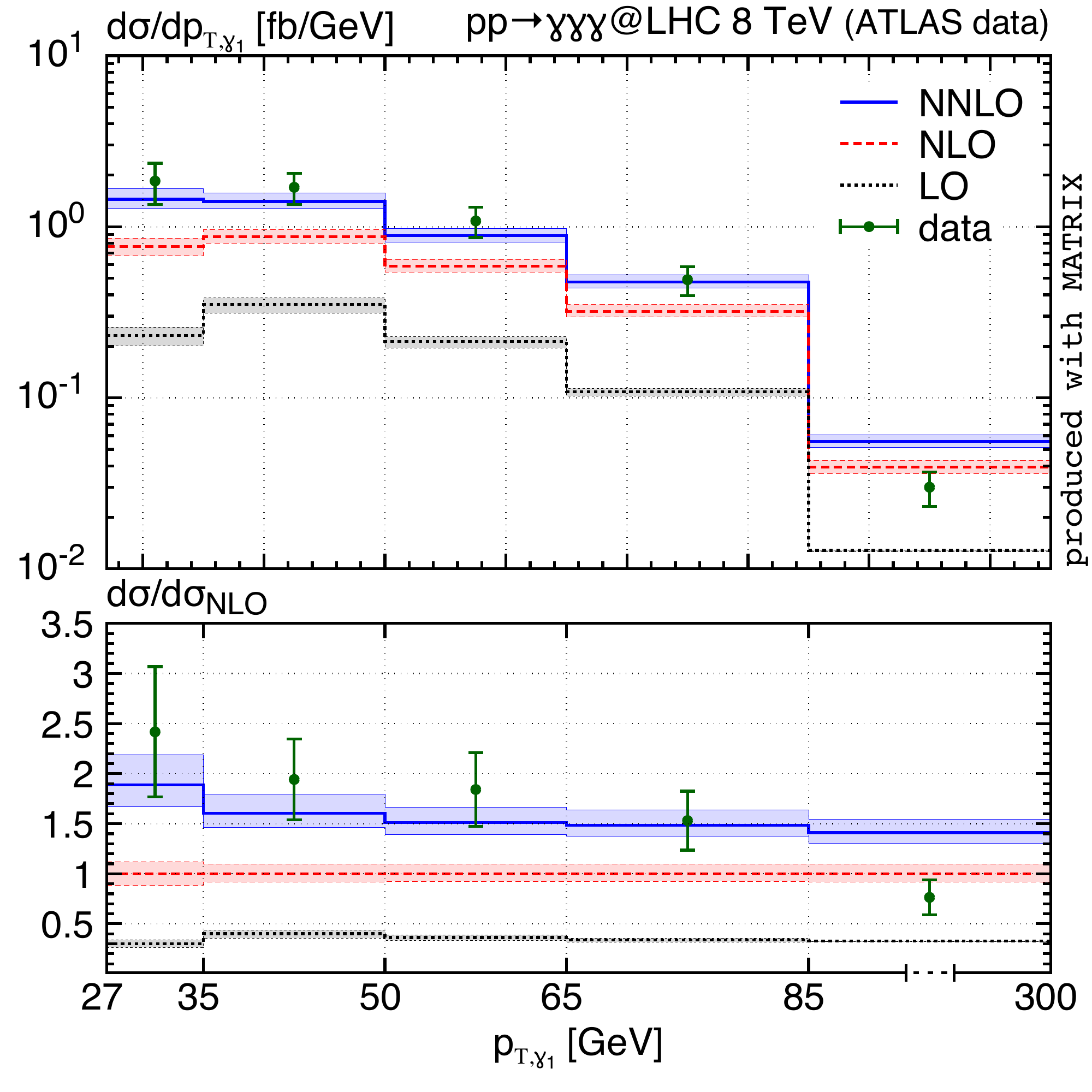} 
&
\includegraphics[width=.42\textwidth]{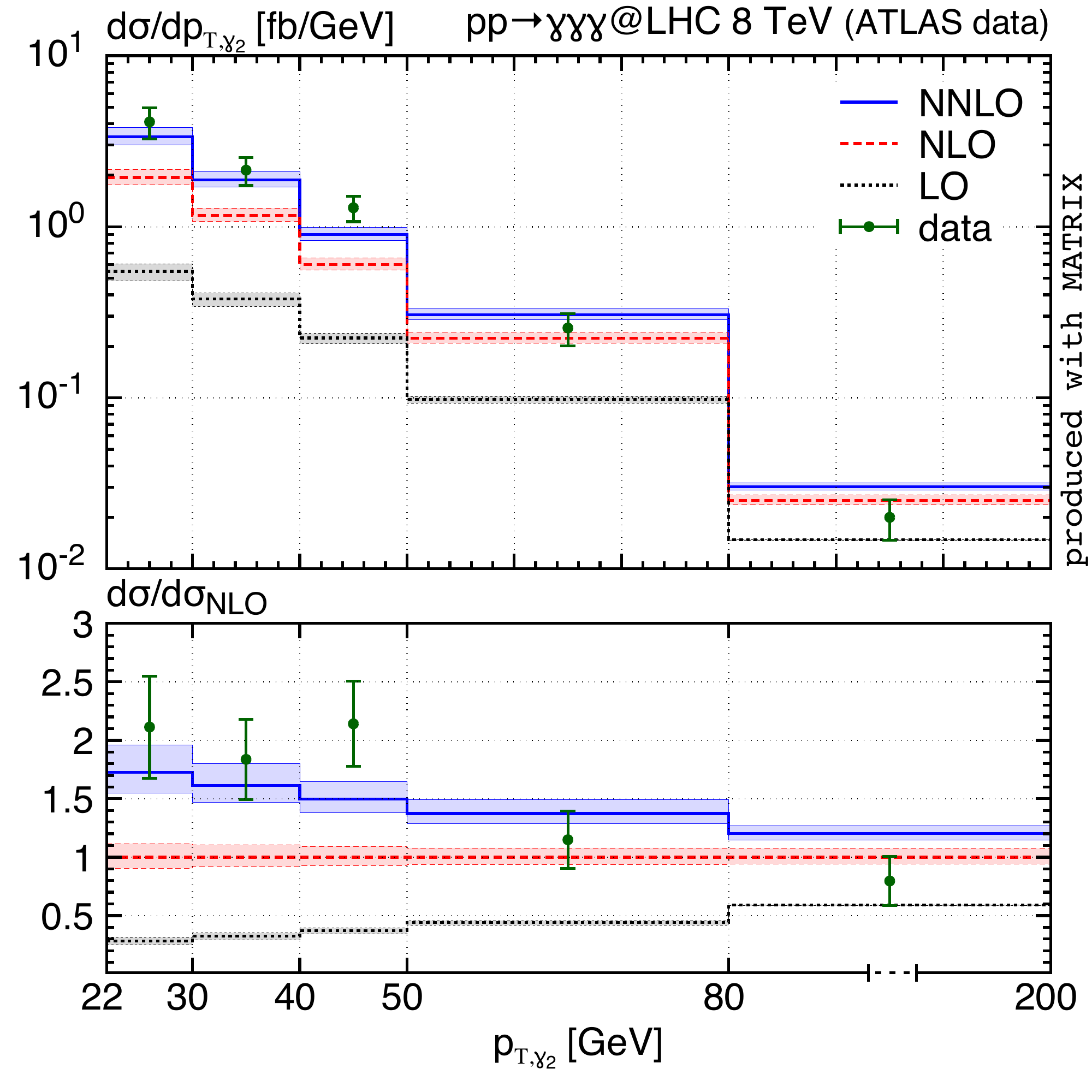}
\end{tabular}
\begin{tabular}{c}
\includegraphics[width=.42\textwidth]{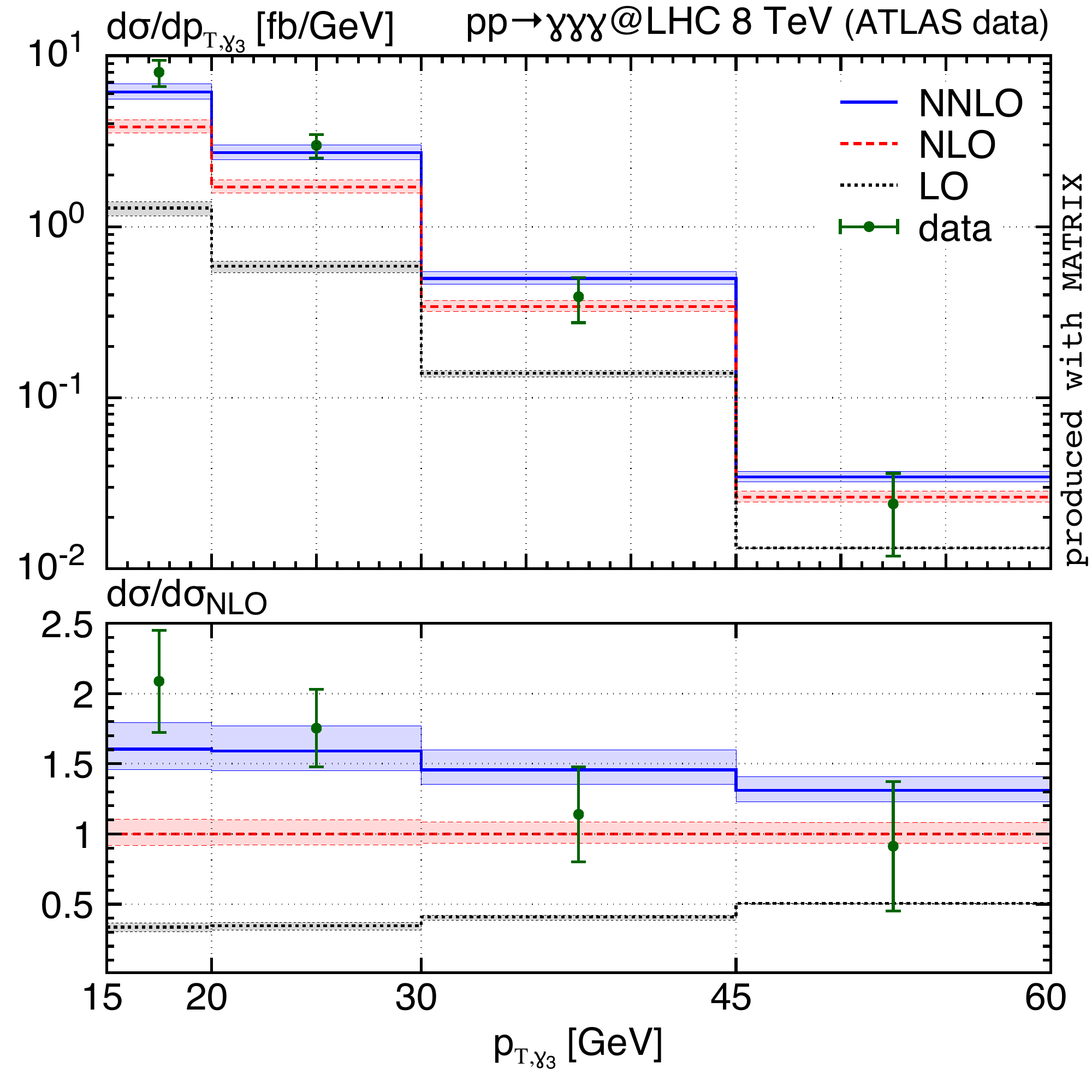}
\end{tabular}
\vspace*{1ex}
\caption{\label{fig:pT} Same as \fig{fig:invmass}, but for the transverse momentum spectrum of each photon.}
\end{center}
\end{figure}

\begin{figure}
\begin{center}
\begin{tabular}{cc}
\includegraphics[width=.42\textwidth]{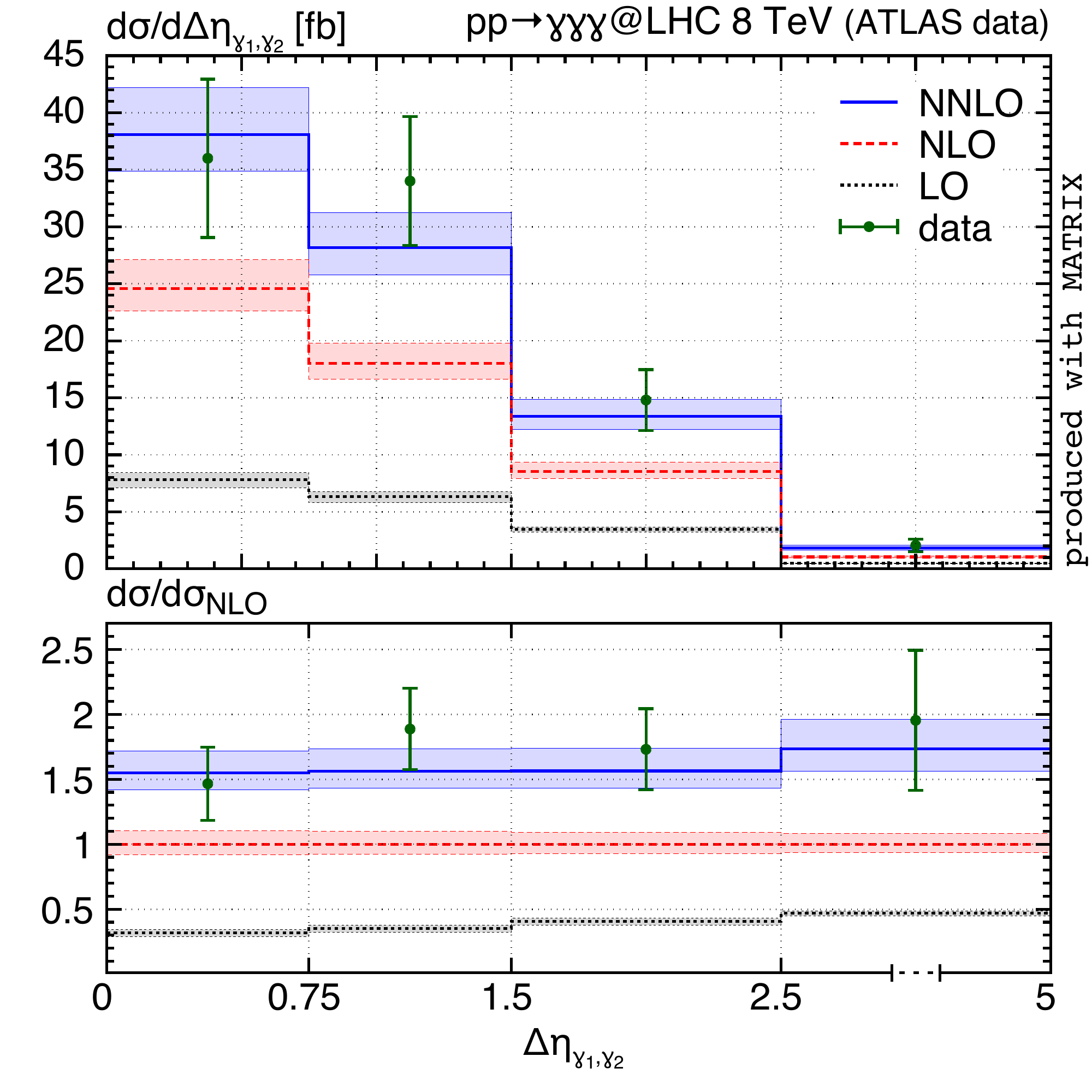} 
&
\includegraphics[width=.42\textwidth]{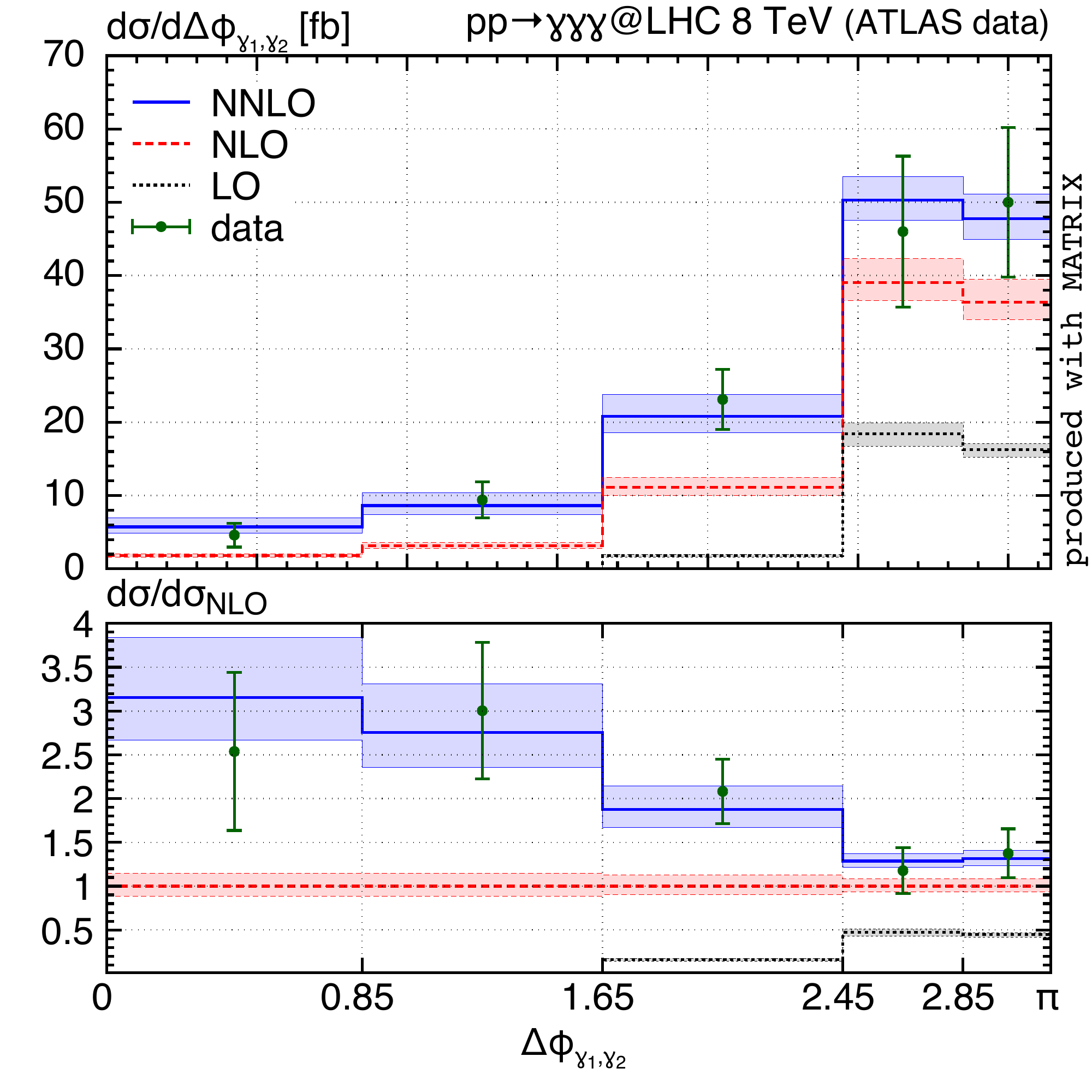}
\end{tabular}
\begin{tabular}{cc}
\includegraphics[width=.42\textwidth]{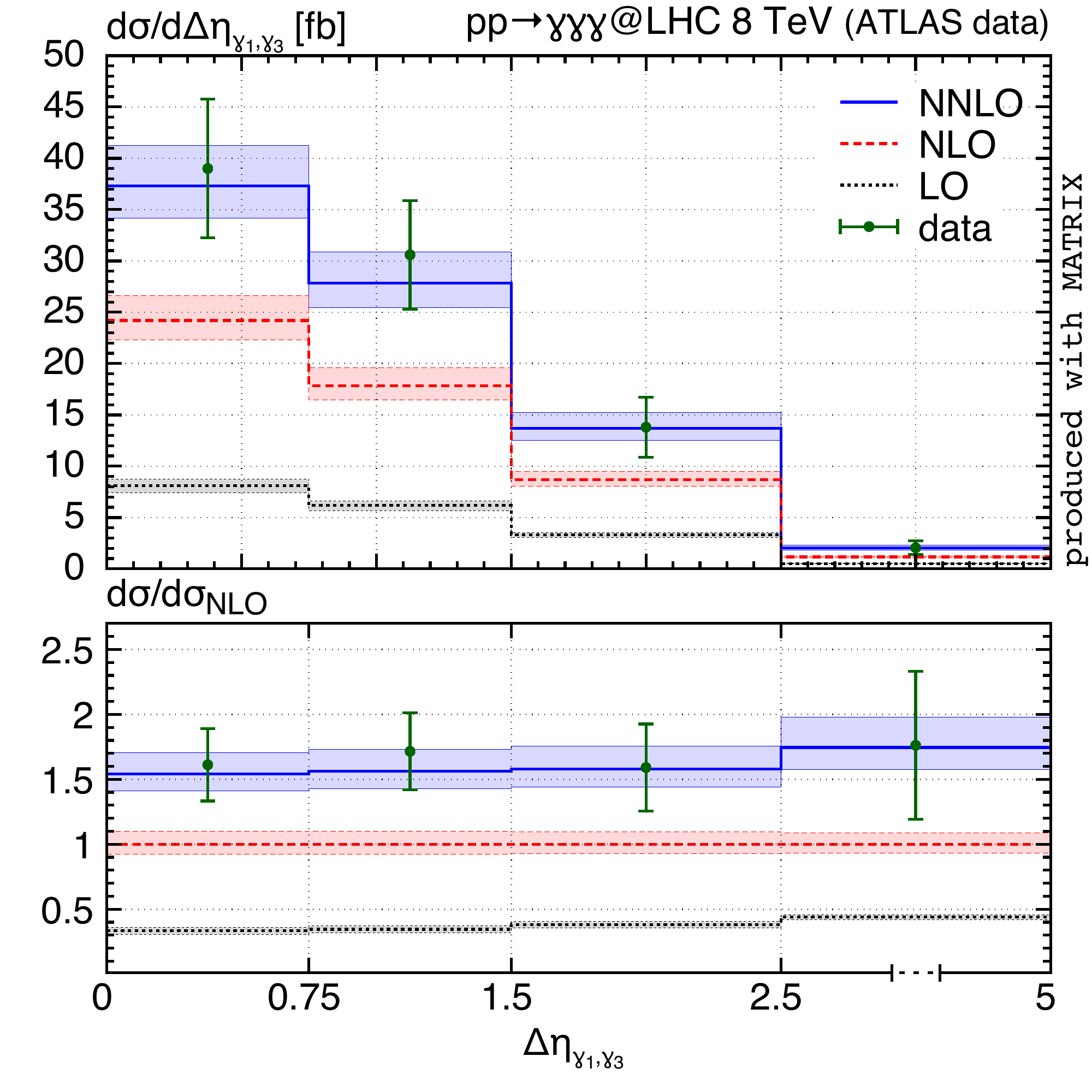}
&
\includegraphics[width=.42\textwidth]{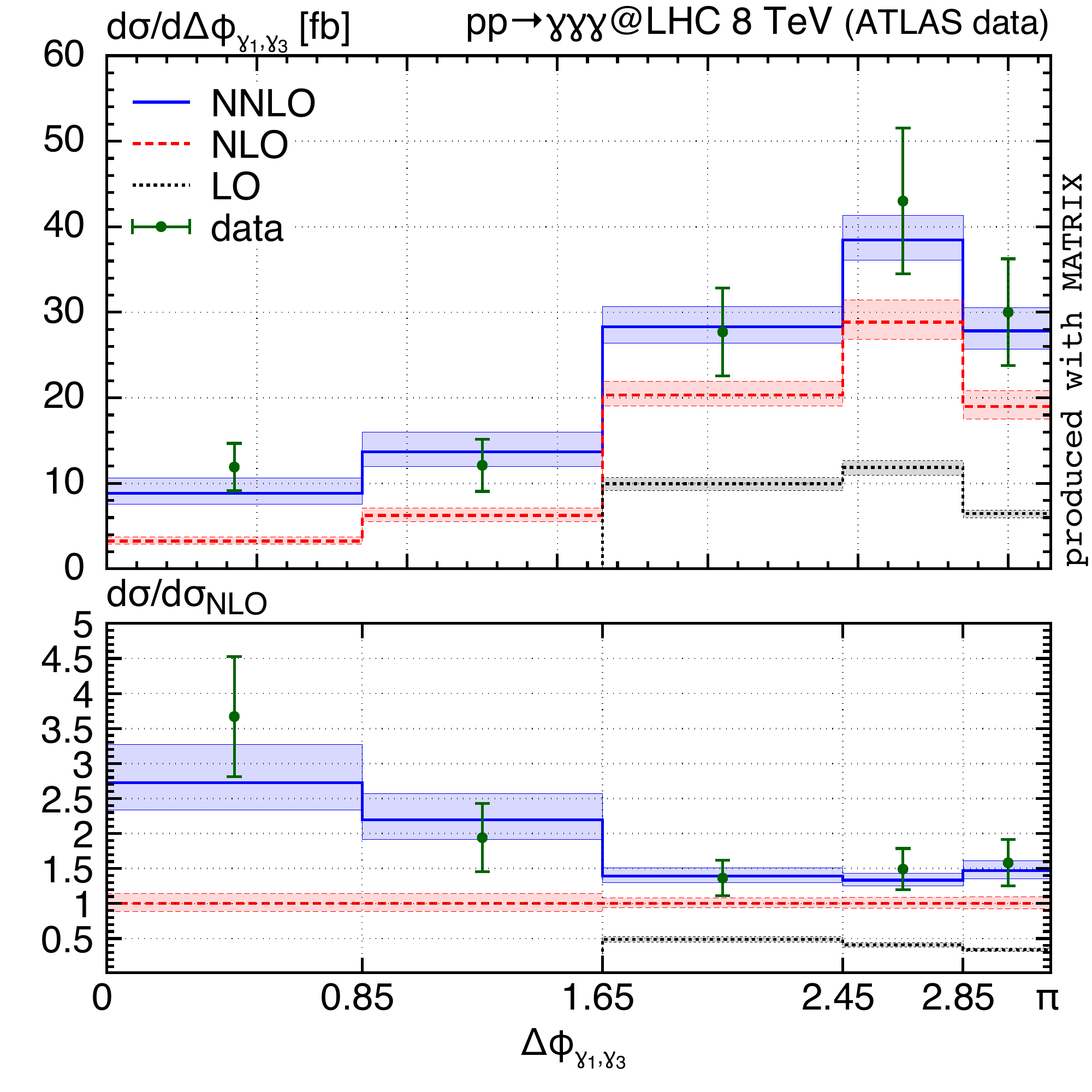}
\end{tabular}
\begin{tabular}{cc}
\includegraphics[width=.42\textwidth]{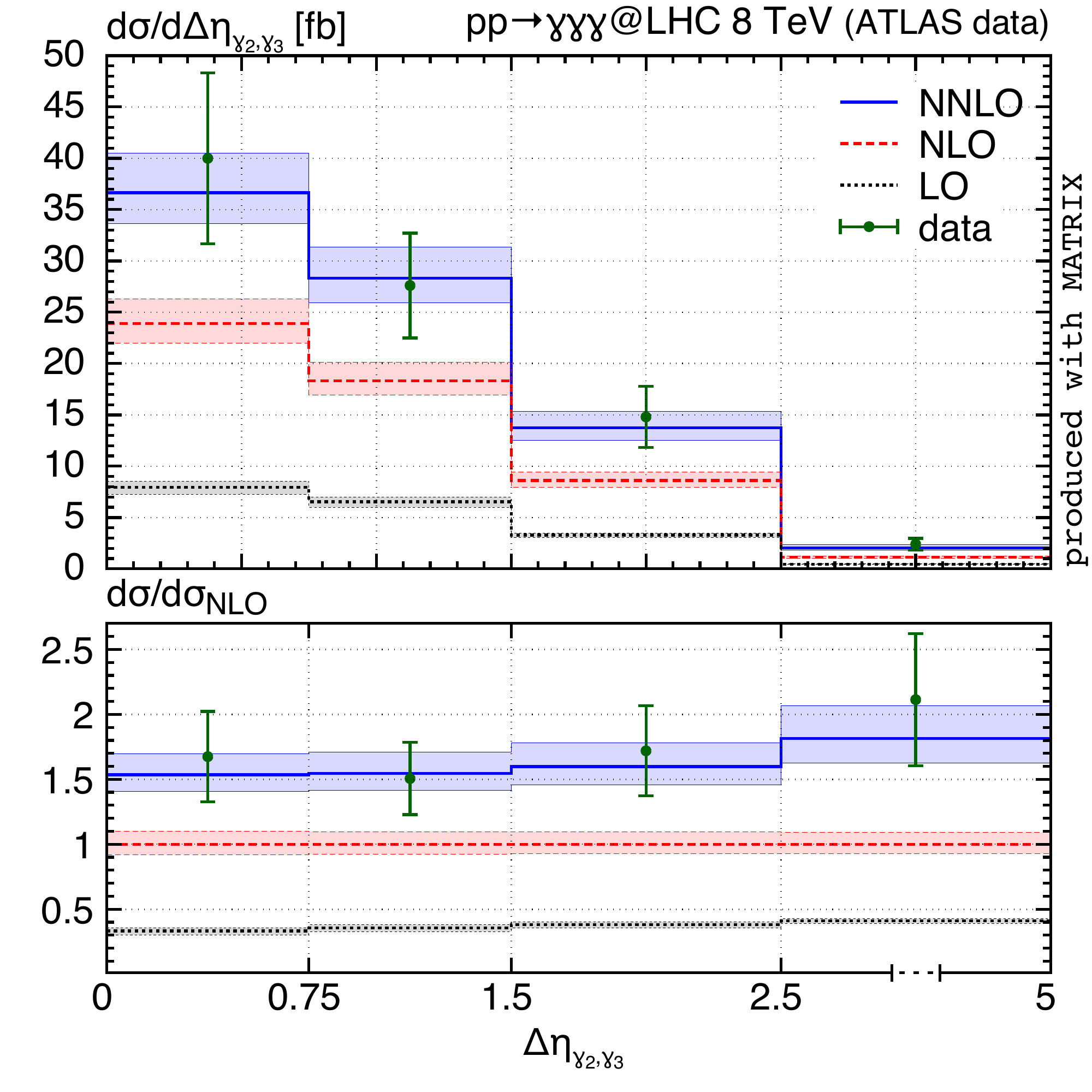}
&
\includegraphics[width=.42\textwidth]{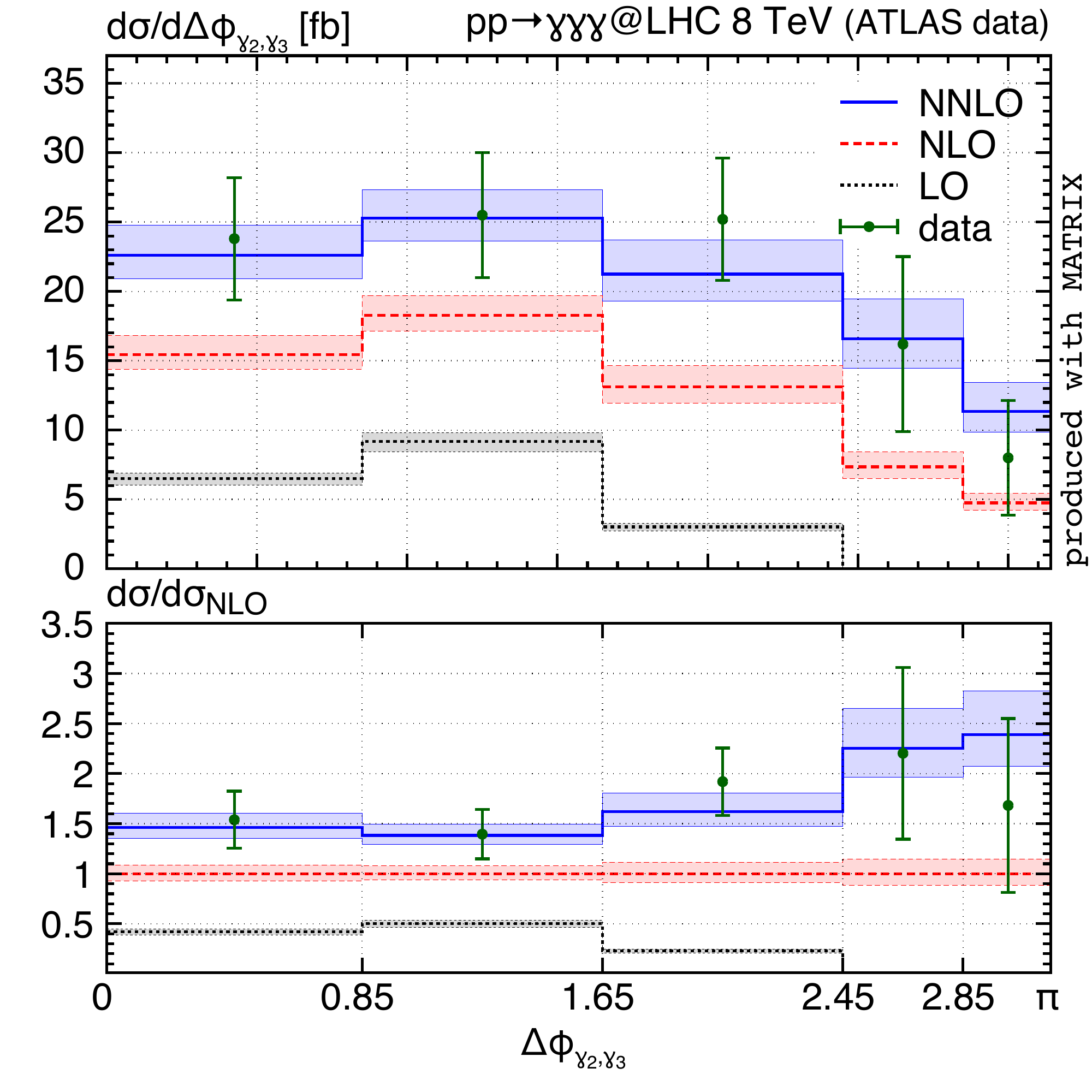}
\end{tabular}
\vspace*{1ex}
\caption{\label{fig:etaphi} Same as \fig{fig:invmass}, but for the difference in $\eta$ and $\phi$ for each photon pair.}
\end{center}
\end{figure}

We continue our discussion by comparing differential distributions at LO, NLO and NNLO to 
ATLAS data at 8\,TeV~\cite{Aaboud:2017lxm} in \figs{fig:invmass}-\ref{fig:etaphi}.
We show the invariant-mass spectrum of the triphoton system (\mggg{}) 
and of photon pairs ($m_{\gamma_i\gamma_j}$) in \fig{fig:invmass},
the transverse-momentum distributions (\ptgi) of the three photons in \fig{fig:pT},
the pseudorapidity differences between two photons ($\Delta\eta_{\gamma_i,\gamma_j}$ with $i,j\in\{1,2,3\}$ and $i\neq j$) in \fig{fig:etaphi}~(left), and  
the difference in the azimuthal angle between two photons ($\Delta\phi_{\gamma_i,\gamma_j}$) 
in \fig{fig:etaphi}~(right).
The agreement between the NNLO predictions and data is truly remarkable.
With only few exceptions, in particular the last bin of the transverse-momentum distributions
of both the hardest ($\ptgone$) and the second-hardest ($\ptgtwo$) photon 
in \fig{fig:pT}, all data points agree with the NNLO predictions
within one standard deviation.
Apart from the pseudorapidity differences, where the NNLO corrections are essentially flat, 
they induce large effects on the shapes of all other distributions. Indeed, besides the corrected normalization these shape distortions at NNLO 
are absolutely crucial to describe the measured distributions well.
In several cases the NNLO/NLO $K$-factor becomes as large as a factor of two, 
and it can even reach a factor of three and more for the $\Delta\phi_{\gamma_i,\gamma_j}$ distributions
in \fig{fig:etaphi}.
Assuming no substantial BSM effects, the excellent agreement with available data suggests that
NNLO scale bands indeed do not substantially underestimate uncertainties due to missing higher-order corrections.
At the same time, it is clear that the LO results including their uncertainties are 
completely insufficient to yield any meaningful prediction, and even NLO predictions 
can hardly be trusted when showing large discrepancies with data of several standard deviations.

\begin{figure}
\begin{center}
\begin{tabular}{cc}
\includegraphics[width=.42\textwidth]{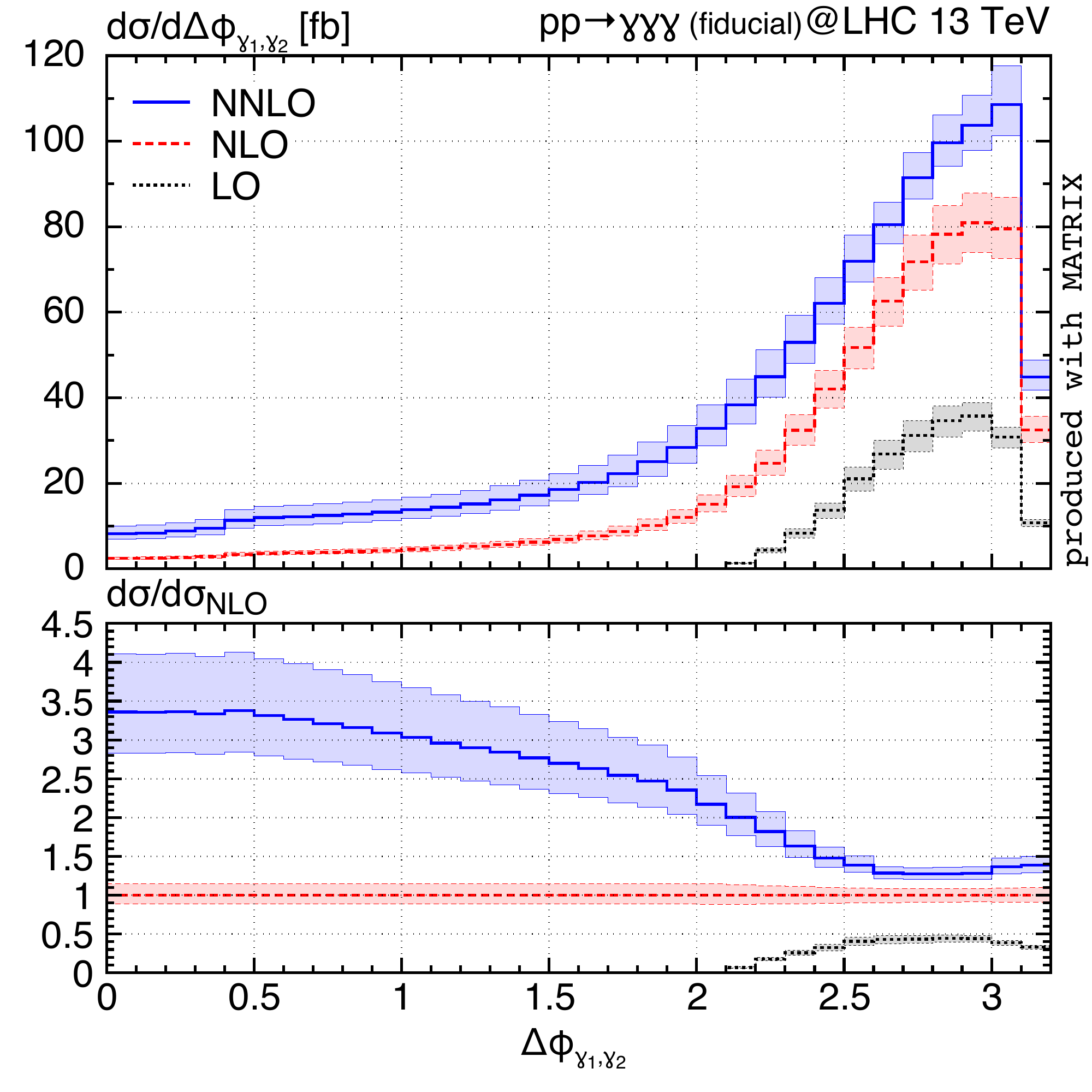} 
&
\includegraphics[width=.42\textwidth]{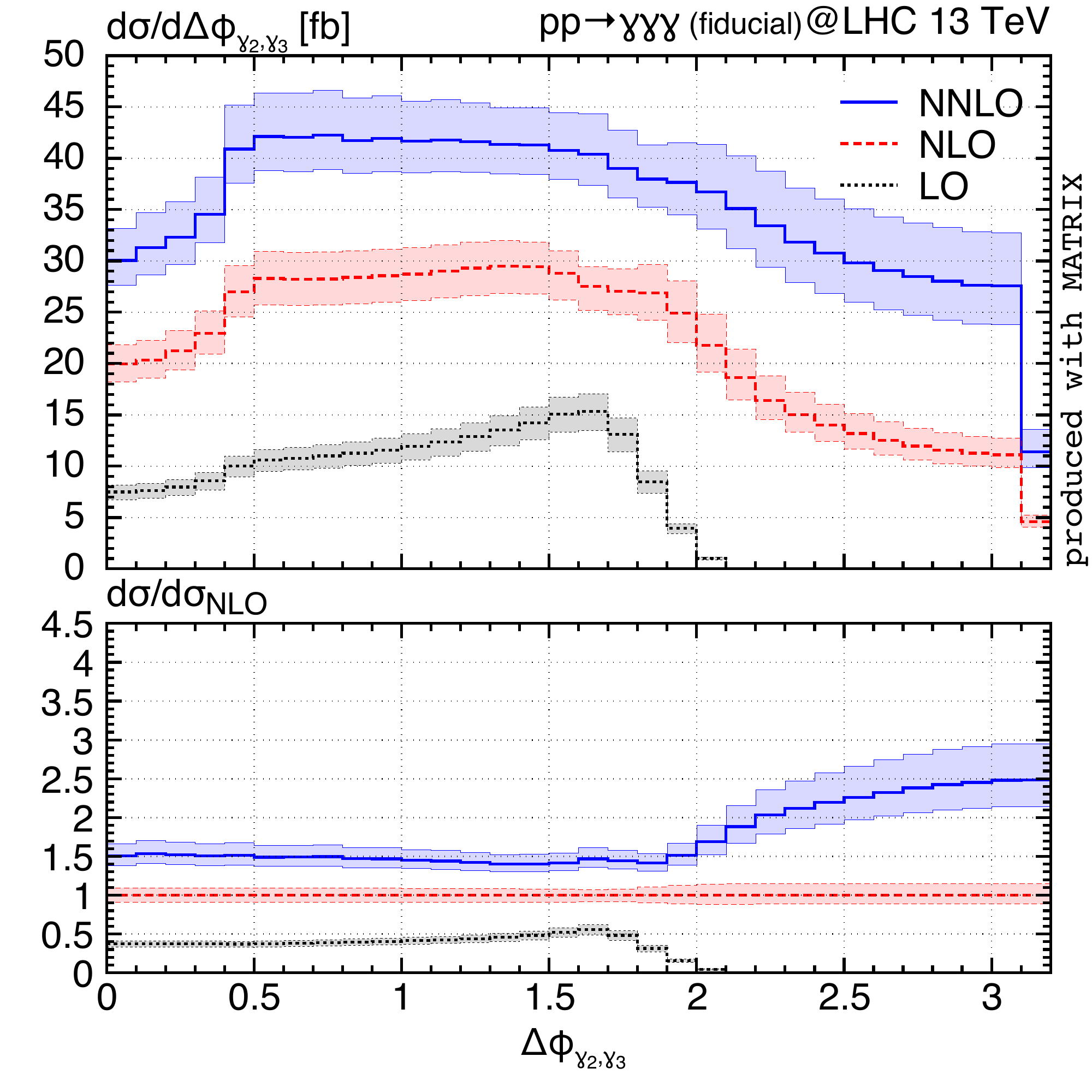}
\end{tabular}
\begin{tabular}{cc}
\includegraphics[width=.42\textwidth]{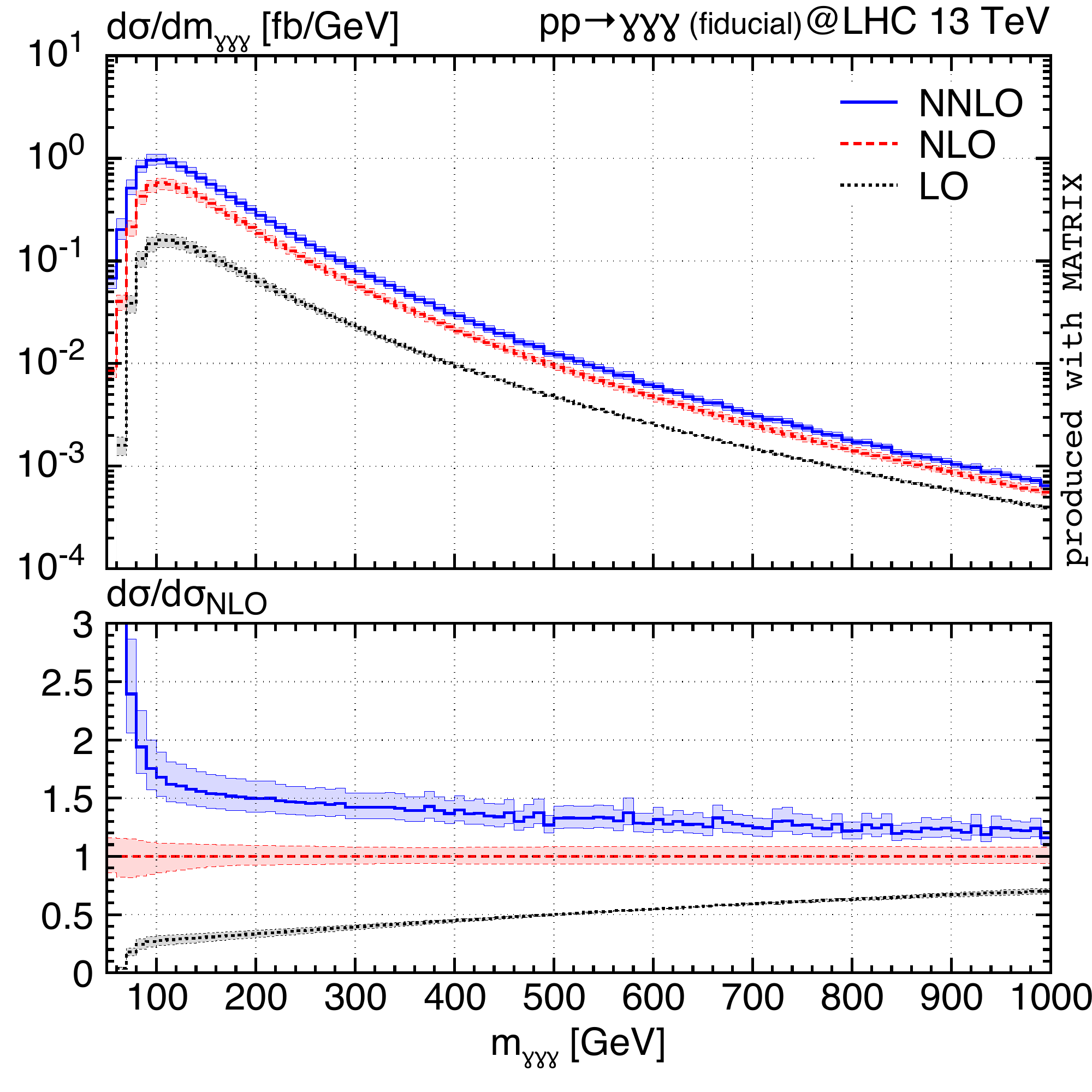}
&
\includegraphics[width=.42\textwidth]{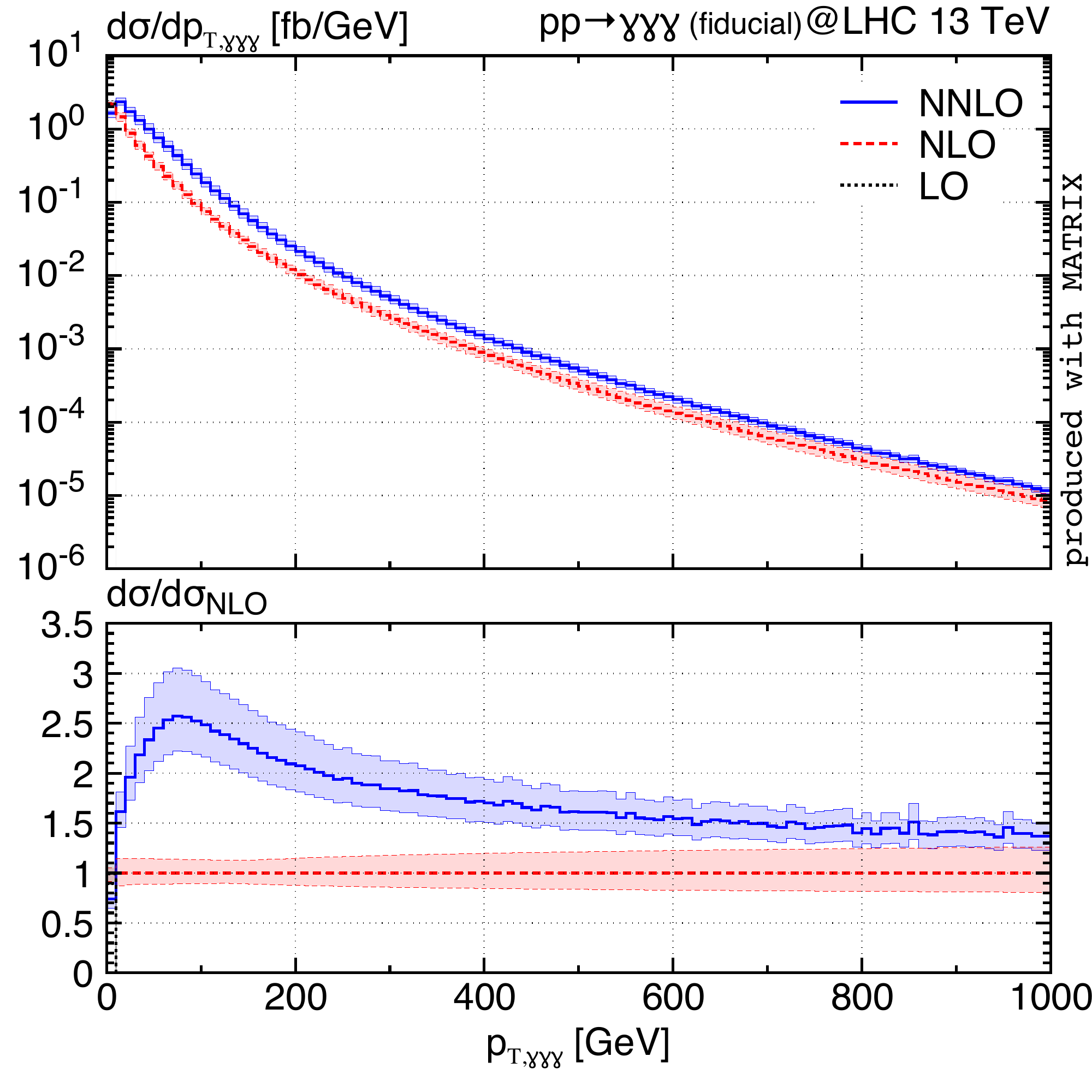}
\end{tabular}
\begin{tabular}{cc}
\includegraphics[width=.42\textwidth]{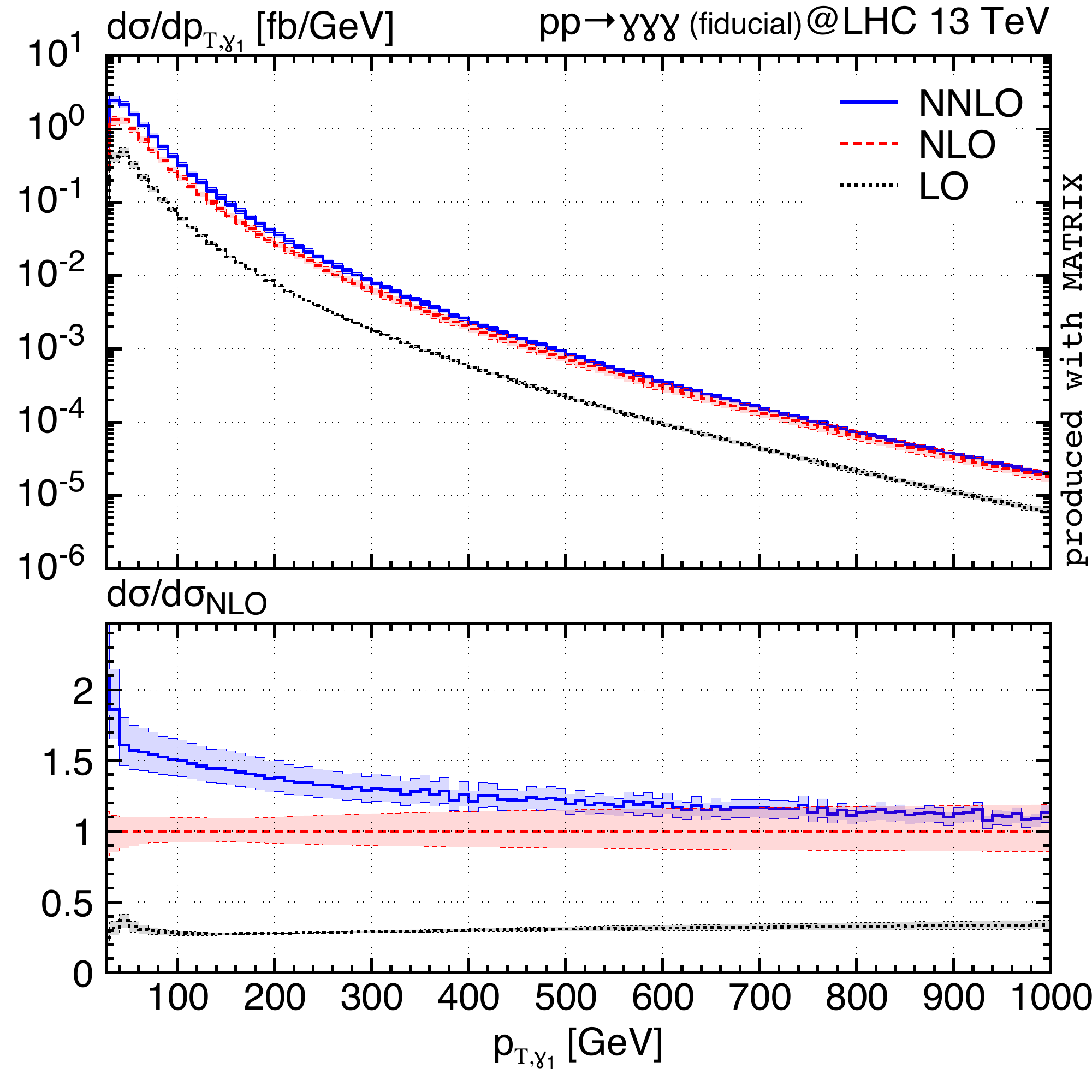}
&
\includegraphics[width=.42\textwidth]{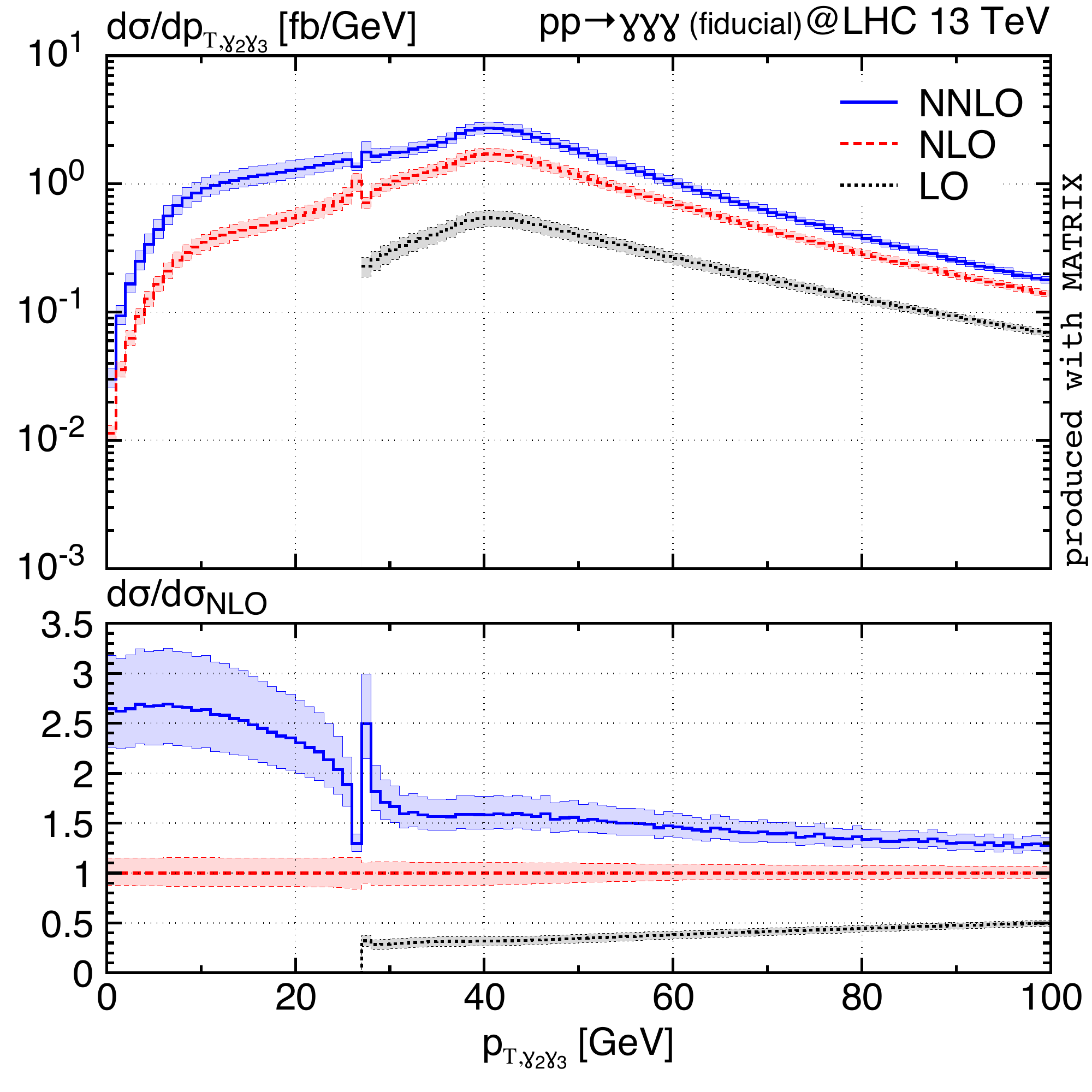}
\end{tabular}
\vspace*{1ex}
\caption{\label{fig:13TeV} Distributions at 13\,TeV without data. The colour coding corresponds to
\fig{fig:cs}.}
\end{center}
\end{figure}

We now move to studying differential distributions and the impact of NNLO corrections for higher centre-of-mass energies. 
Unfortunately, there has not yet been any 13\,TeV measurement at the LHC, which is why we focus on theoretical predictions only.
We obtain differential results for all collision energies considered in \tab{tab:cs}.
In comparison to the results at $\sqrt{s}=8$\,TeV, we do not find any substantial differences in the shapes of the $K$-factors.
The corrections generally increase with energy, which we have already inferred from the results in \tab{tab:cs}, and, as expected, the 
jet activity increases. Since our results are not NNLO accurate any longer when requiring a jet, we focus on distributions that involve only the kinematics of the colour singlet final state. We note, however, that some observables intrinsically require jet activity in certain phase space regions, which is reflected by vanishing LO predictions.
All relevant features will be discussed using the $13$\,TeV results as a reference.

In \fig{fig:13TeV} we present various differential distributions at 13\,TeV,
and we use a much finer binning to better resolve certain features.
The upper plots of \fig{fig:13TeV} show the azimuthal difference between the hardest and
the second-hardest photon ($\Delta\phi_{\gamma_1,\gamma_2}$) as well as between the second-hardest
and the third-hardest photon ($\Delta\phi_{\gamma_2,\gamma_3}$).
The hierarchy of the $p_T$-ordered photons
induces significant differences between the two cases.
For LO kinematics, $\gamma_1$ and $\gamma_2$ need to recoil against each
other since $\gamma_3$ does not carry sufficient energy to provide the recoil when the momenta of
those two harder photons align.
Correspondingly, $\gamma_2$ and $\gamma_3$ cannot be produced in back-to-back configurations at LO since these two photons need to recoil against the hardest photon $\gamma_1$.
As a consequence, the LO cross section vanishes for $\Delta\phi_{\gamma_1,\gamma_2}<2\pi/3$
and $\Delta\phi_{\gamma_2,\gamma_3}>2\pi/3$, respectively.
Those phase space regions are filled only upon inclusion of
real QCD radiation through higher-order corrections, which
is required to overcome the kinematic constraints at LO.
Accordingly, the NLO (NNLO) predictions in these regimes are effectively only LO (NLO) accurate,
which is reflected by the increased size of both corrections and uncertainty bands.
We find that back-to-back configurations of $\gamma_1$ and $\gamma_2$
are still preferred at higher orders,
whereas the distribution of the azimuthal separation between $\gamma_2$ and $\gamma_3$
becomes much more uniform when adding higher-order corrections.

In the central plots of \fig{fig:13TeV} we show the invariant-mass and 
transverse-momentum distributions of the three-photon system. The invariant-mass
distribution peaks around $100$\,GeV. Below the peak the distribution 
falls off steeply with a lower bound imposed by the
phase space selection cut $\mggg\geq50$\,GeV.
In that low $\mggg{}$ region radiative corrections increase quite strongly. 
By contrast, higher-order corrections become successively smaller in the tail 
of the $\mggg$ distribution, which is an important region for new-physics 
searches through small deviations from SM predictions. 
Around $\mggg=1$\,TeV NLO and NNLO predictions become almost compatible 
within uncertainties. Also in the tail of the triphoton transverse-momentum 
spectrum NNLO corrections become successively smaller. We note that this 
observable vanishes for $\ptggg>0$ at LO, and it diverges for $\ptggg\rightarrow 0$ at any order 
in QCD perturbation theory. Only a suitable resummation of logarithmic 
contributions (cf.~\citere{Kallweit:2020gva}) warrants a physical prediction at small $\ptggg$.
The increase of NLO (NNLO) uncertainties in the $\ptggg$ spectrum is again
caused by the fact that the effective accuracy is decreased to LO (NLO)
for this observable.

We conclude our analysis by studying the transverse-momentum spectrum 
of the hardest photon ($\ptgone$) and of the sum of the 
second- and third-hardest photons ($\ptgtwothree$) in the lower plots of \fig{fig:13TeV}. 
The distribution in $\ptgone$ shows a rather similar pattern as the $\mggg{}$ distribution in terms 
of the NNLO corrections. The distribution is cut at $\ptgone=27$\,GeV in our fiducial setup,
which induces an interesting behaviour in the $\ptgtwothree$ spectrum around that value at NLO and NNLO.
The reason is that $\ptgtwothree=\ptgone$ for LO kinematics, so that this distribution is not filled below $27$\,GeV at LO.
At higher orders the spectrum develops a perturbative instability 
at this threshold caused by an incomplete 
cancellation of virtual and real contributions from soft gluons, which
is logarithmically divergent, but integrable~\cite{Catani:1997xc}. This instability slightly decreases 
when going from NLO to NNLO, but only a resummation of the relevant 
logarithmic contributions would yield a stable result.
Practically, choosing a wide bin around 
the instability would alleviate this unphysical behaviour significantly. 
Note that such perturbative instabilities are present also for the $\ptgonetwo$ and
$\ptgonethree$ spectra, at the fiducial cut values imposed on $\ptgthree$ and $\ptgtwo$, respectively.

We stress again that LO results clearly fail to provide any reasonable 
prediction, and even NLO results strongly underestimate the cross 
section. Thus, one should bear in mind that care must be taken when determining new-physics
contributions through higher-dimensional operators, see \citeres{Aad:2015bua,Denizli:2019ijf,Denizli:2020wvn} for instance. 
Relying on LO cross sections for that matter might be insufficient because of the vastly 
inappropriate modelling of the cross section and distributions at this order.
Even when relative BSM effects are computed by taking the ratio to the 
SM prediction at the same order, those effects are hardly trustworthy when 
based on a calculation at LO in QCD.

To summarize, we have presented a new calculation of the NNLO QCD corrections to the hadronic production of 
three isolated photons. This is the very first $2\to 3$ process known at this accuracy, and we found our 
results to be fully compatible with the earlier calculation of \citere{Chawdhry:2019bji}. Our calculation shows 
that even for highly non-trivial colour singlet processes \qt{} subtraction is suitable to obtain NNLO predictions
with systematic uncertainties at the few permille level, and that even complex five-point two-loop amplitudes can be 
implemented so efficiently that they do not pose a numerical bottleneck and can be evaluated directly during phase space integration.
We have presented numerical results at various $pp$ collision energies. Radiative corrections to this process
are enormous: The NLO/LO $K$-factor is a factor of three (at $7$\,TeV) to five (at $100$\,TeV), while NNLO 
corrections turn out to be still as large as $+50$\% (at $7$\,TeV) to $+100$\% (at $100$\,TeV).
NNLO is the first order to yield reliable predictions for triphoton production. Indeed, the comparison to
8\,TeV data indicates substantial discrepancies with NLO results, while being in excellent agreement with 
the NNLO predictions. This is true not only for the fiducial rate, but also for differential distributions in the fiducial 
phase space. Our study of distributions at $13$\,TeV substantiates that NNLO accuracy is mandatory 
for precision phenomenology of this process in the future. We reckon that our results and our calculation\footnote{The 
implementation of NNLO corrections to triphoton production will be made publicly available with the next release of \Matrix{}. A 
preliminary version of the code is available from the authors upon request.}
will be very useful for both measurements of the production of three isolated photons at the LHC and related BSM searches.

\noindent {\bf Acknowledgements.}
We are indebted to Fernando Febres Cordero, Massimiliano Grazzini, and Giulia Zanderighi for fruitful discussions and 
comments on the manuscript. 
We thank Josu Cantero Garcia and Kristin Lohwasser for correspondence and clarifications regarding the 
ATLAS 8\,TeV results~\cite{Aaboud:2017lxm}. Furthermore, we are grateful to Micha\l{} Czakon and Rene Poncelet for correspondence 
about the results in the ancillary files of \citere{Chawdhry:2019bji}. Finally, we would like to thank Marco Bonvini for useful discussions 
on the alternative approach to estimate theoretical uncertainties~\cite{Bonvini:2020xeo}.
The work of SK is supported by the ERC Starting Grant 714788 REINVENT.
The work of VS is supported by the European Research Council (ERC) under the European Union's Horizon 2020 research and innovation programme,
\textit{Novel structures in scattering amplitudes} (grant agreement No.\ 725110).
\setlength{\bibsep}{3.1pt}
\renewcommand{\em}{}
\bibliographystyle{apsrev4-1}
\bibliography{triphoton}
\end{document}